\documentclass[article]{revtex4-1} 
\usepackage[T1]{fontenc}
\usepackage[utf8]{inputenc}
\usepackage[english]{babel}
\usepackage{amsmath,amssymb,amsfonts}
\usepackage{graphicx,graphics}
\usepackage{subfigure}
 \usepackage{caption}
\usepackage{multirow}
 \usepackage{dcolumn}
\usepackage{bm}
\usepackage{color}
\usepackage{amsthm}             
\usepackage{amssymb}
\usepackage{mathrsfs}
\usepackage{dcolumn}
\usepackage{bm}
\usepackage{epsfig}
\usepackage{epstopdf}
\usepackage{float}
\usepackage{multirow}

\begin{document}

\title{On the electrostatic interactions involving long-range Rydberg molecules}
\author{H. Rivera-Rodr\'{\i}guez}\email{hrivera@ciencias.unam.mx}
\affiliation{Instituto de F\'{\i}sica, Universidad Nacional Aut\'onoma de M\'exico, Apdo. Postal 20-364, 01000 Cd. de M\'exico, M\'exico}

\author{R. J\'auregui}\email{rocio@fisica.unam.mx}
\affiliation{Instituto de F\'{\i}sica, Universidad Nacional Aut\'onoma de M\'exico, Apdo. Postal 20-364, 01000 Cd. de M\'exico, M\'exico}

\begin{abstract}
A ground state atom immersed in the wave function of the valence electron of a Rydberg atom can generate a long-range Rydberg molecule (LRRM).
In this work,  using  the multipole expansion of the electrostatic interaction in prolate spheroidal coordinates, approximate and compact expressions of the electrostatic potential that determine the chemistry of trilobite and butterfly LRRM are explored. It is shown that even the spheroidal monopole term can be used to describe general features of the potential
generated by a LRRM at short distances.  It is also shown that even at long separations that allow a perturbative description of  the intermolecular interaction between two LRRM, the convergence of the multipole spheroidal expansion is faster than that of its spherical analogue.

\end{abstract}

\maketitle

Long-range Rydberg molecules (LRRM) result from an attractive interaction of the
highly excited electron of a Rydberg atom colliding with one or more than one ground-state atoms. 
They provide a unique scenario for the study of both Rydberg and ultra cold scattering physics.
LRRM general features can be properly described in terms of a novel chemical binding mechanism: the Fermi pseudo potential model \cite{Fermi1934} for s-wave \cite{green2000} and p-wave \cite{khuskivadze2002,hamilton2002} scattering. These molecules were predicted to have huge tunable bond lengths --similar to the size of the Rydberg atoms-- and weak binding energies. As the electron is extremely localized near the perturber atom, the resulting charge separation yields permanent electric dipole moments even for  homonuclear LRRM.  These amazing properties have already been verified  in Rb \cite{pohl2011,bellos2013, krump2014,anderson2014, niederprum2016}, Cs \cite{tallant2012,booth2015} and Sr \cite{desalvo2015}. Heteronuclear \cite{peper2021} and ion-Rydberg atom \cite{duspayev2021} LRRM were reported recently. The experiments also provide sound advances on the engineering of the quantum chemistry of LRRM by the control of their bond length, vibrational state, angular momentum and orientation in a weak external electric field \cite{niederprum2016}.

Trilobite and butterfly molecules are  diatomic  LRRM  derived from the s-wave and p-wave scattering  respectively. Their electron wave function maximizes the contribution of the corresponding contact interaction. As a consequence, the resulting orbital mixes all high angular momentum states of the Rydberg electron, and the corresponding electron density acquires a characteristic distribution directly related to their names, Fig.~1. The set of dimer electron orbitals is a basis  for the study of LRRM  under different environmental conditions and the theoretical analysis of long range polyatomic molecules \cite{liu2006,liu2009,eiles2016}. The natural coordinate system for their description corresponds to prolate spheroidal  coordinates \cite{gay1989,granger2001}. A compact approximate expression of such orbitals based on Coulomb Green’s function formalism is known  \cite{blinder1993,bartell1996,chibisov2000}.
 It encodes the general attributes of trilobite and butterfly electronic states, though accurate spectroscopic predictions require more complex formulations, in particular those incorporating relativistic effects \cite{kleinbach2017,eiles2017}.

The permanent spherical dipole moment of a diatomic LRRM can be enormous. However, both higher spherical multipole moments and their fluctuations are not negligible. As a consequence the quantum chemistry of these molecules requires a description beyond the standard dipole physics expectations. Since the prolate spheroidal  coordinates provide the inherent  system to describe the morphology  of trilobite and butterfly electronic states, here we perform an alternative analysis of the electrostatic interactions  based on a multipole expansion in these coordinates.  It is found that even the monopole spheroidal potential displays the general features of the electrostatic potential of an isolated LRRM.  By exploring the long range interactions between trilobite and butterfly  molecules, this study reveals interesting properties of simple polyatomic LRRMs and the dynamics behind  their conformation.

\section{ LRRM orbitals.}
\begin{figure}
\centering
\begin{tabular}{@{}c @{}c @{}c}
\includegraphics[width=.33\linewidth]{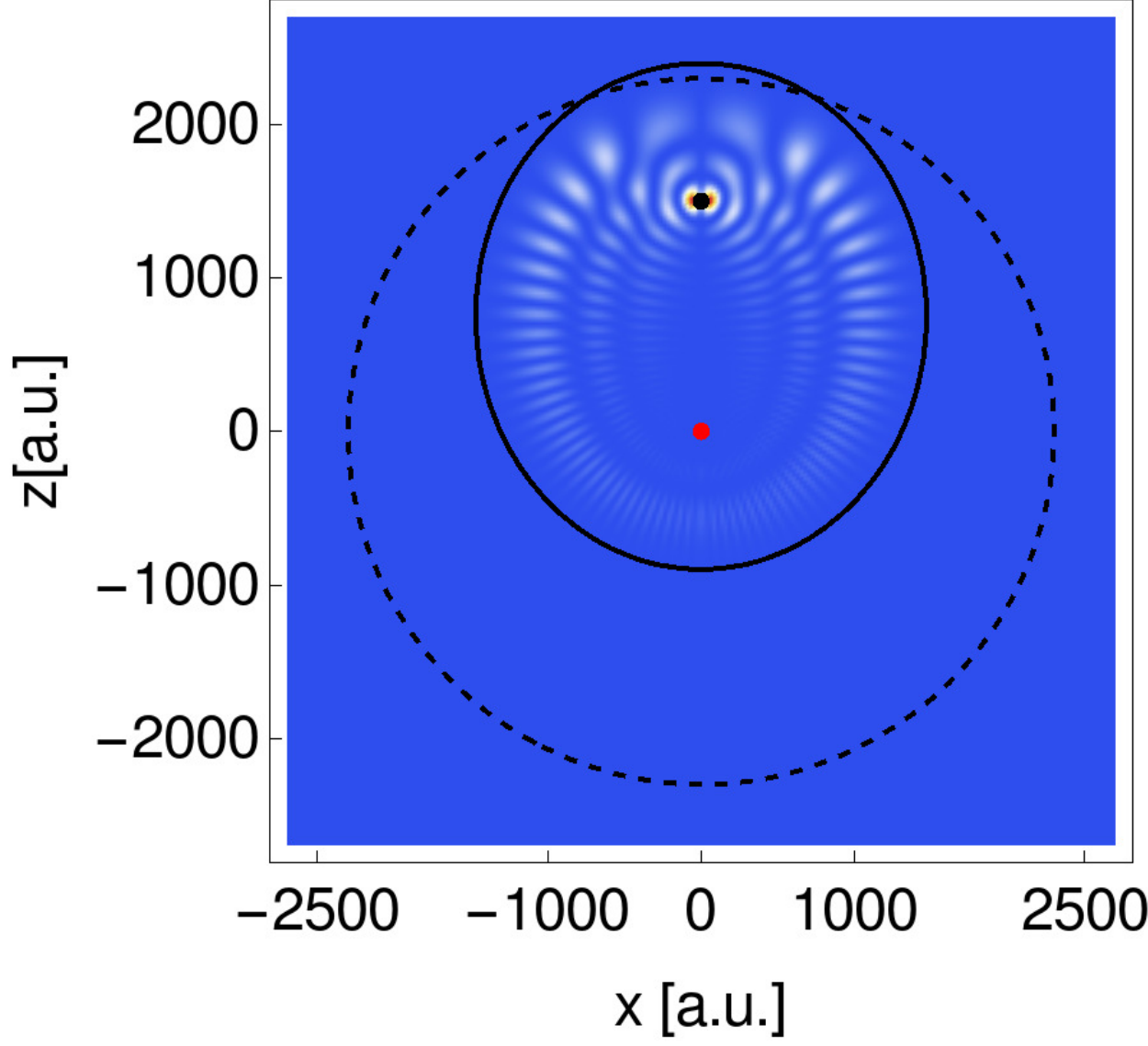}&
\includegraphics[width=.33\linewidth]{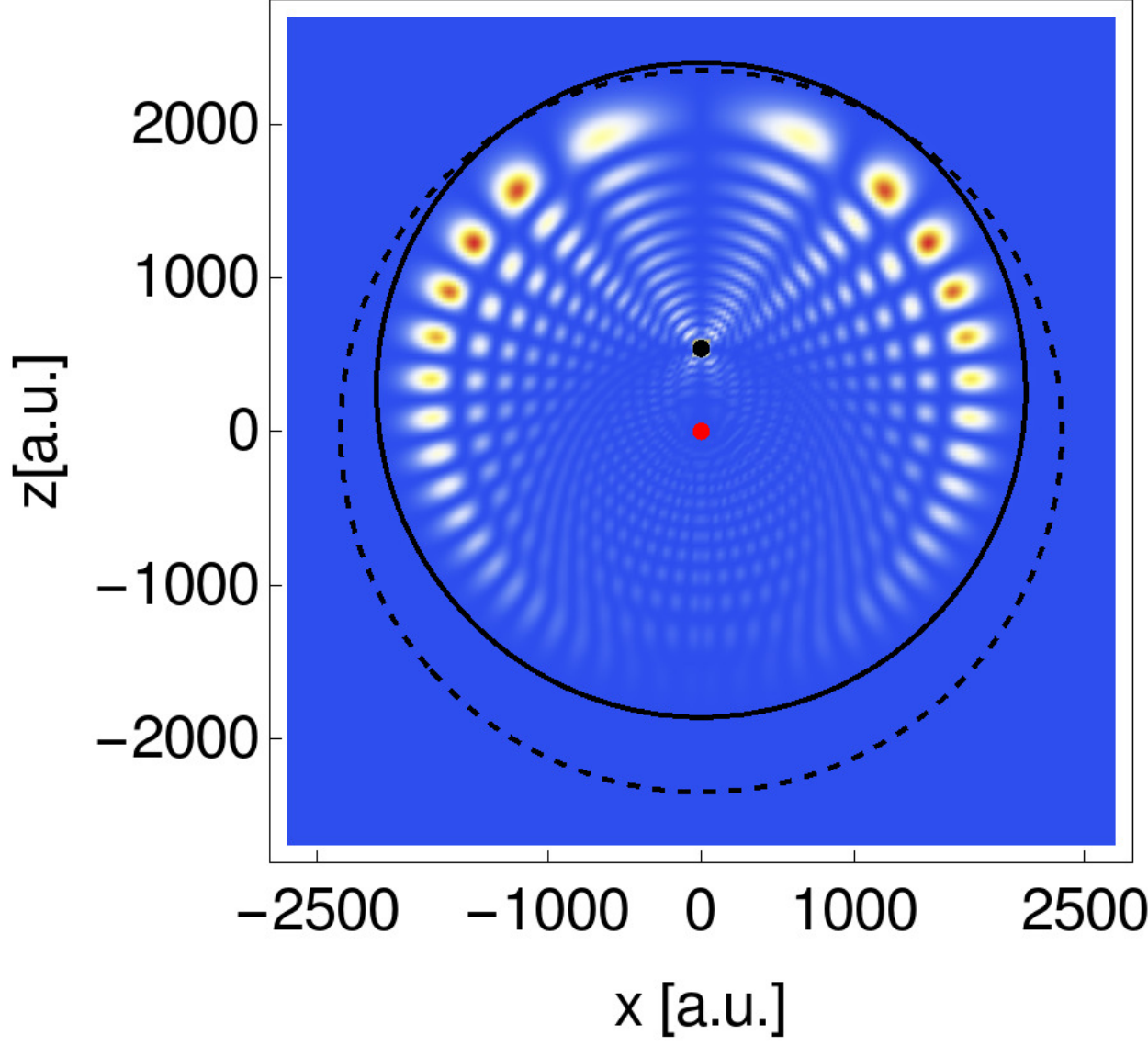}&
\includegraphics[width=.33\linewidth]{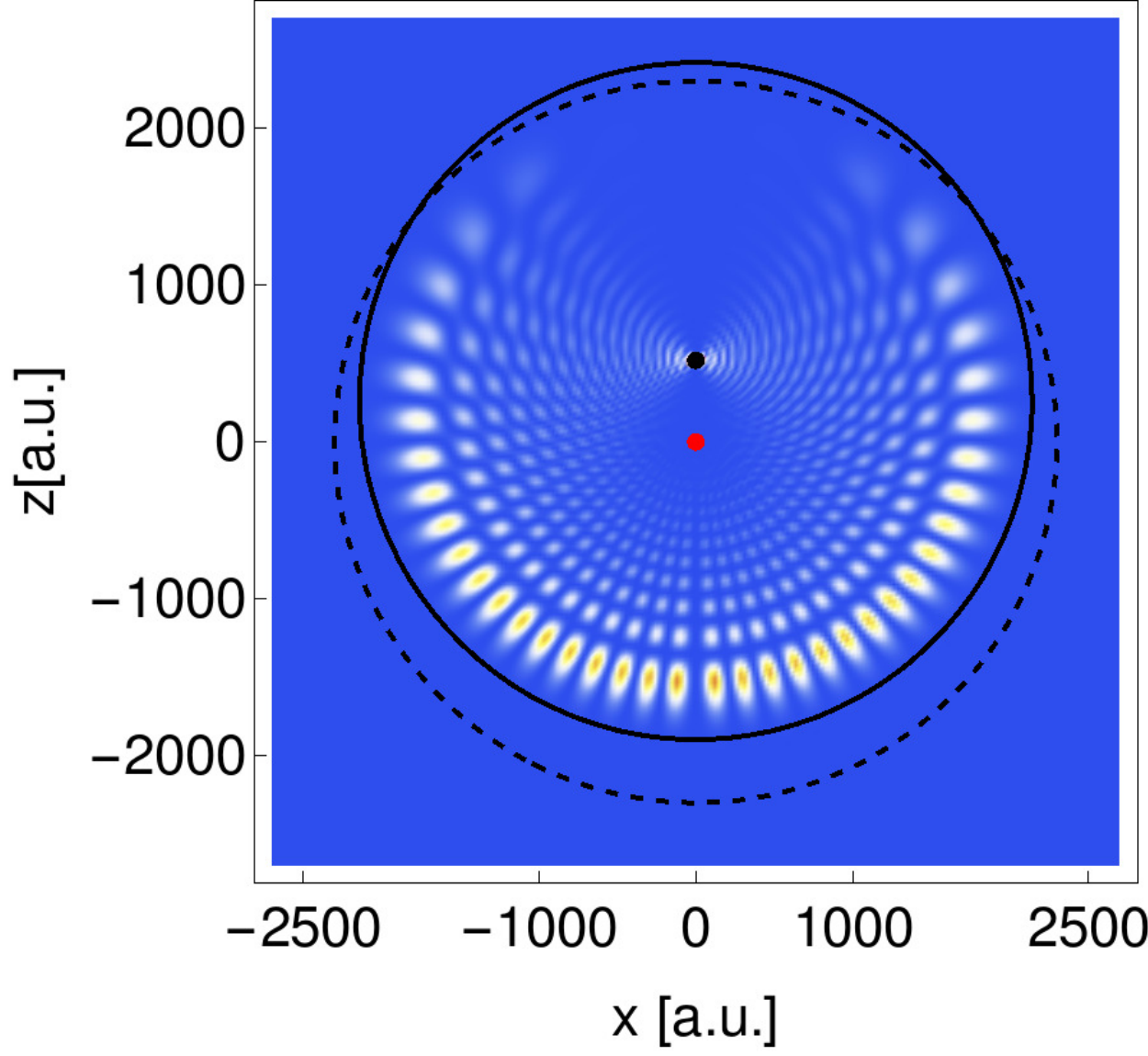}\\
(a) &(b)& (c)\\
\end{tabular}
\caption{ Electronic probability density in the $xz$ plane for Rb molecules with $n=33$.
The Rydberg core is represented by the red dot and the neutral atom by the black one. (a) A trilobite with $\mathsf{r}_\mathrm{t}= 1500$ a$_0$, (b)
a radial butterfly with $\mathsf{r}_{\mathrm{rb}}=540$ a$_0$ and (c) an angular butterfly with $\mathsf{r}_{\mathrm{ab}}=520$ a$_0$.  An ellipsoid $\xi_0$ (solid) and a sphere $r_0$ (dotted) lines in the plane are illustrated, for (a) $\xi_0=2.2$ and $r_0=1.53 \mathsf{r}_{\mathrm t}$, for (b) $\xi_0=7.9$ and $r_0=4.26 \mathsf{r}_{\mathrm{rb}}$, and for (c) $\xi_0=8.3$ and $r_0=4.52 \mathsf{r}_{\mathrm{ab}}$. Length is measured in units of the Bohr radius a$_0$.}	\label{fig:regiones}
\end{figure}

 Since the Rydberg valence electron  has both its ion core  position C and the neutral atom position N as its two natural centers of motion,   prolate spheroidal coordinates  are adequate for the description of the electron orbitals. Let $P$ be a point in space, if $P$ is at distances $r_{PC}$ and $ r_{PN}$ from the two centers $C$ and $N$ respectively,  the spheroidal coordinates $(\xi, \eta, \varphi)$ of $P$ are defined as
\begin{equation}
\xi=\frac{r_{PC}+r_{PN}}{\mathsf{r}}, \quad \eta=\frac{r_{PC}-r_{PN}}{\mathsf{r}}
\label{eq:def}
\end{equation}
where $\mathsf{r}$ is the separation between the two centers; the angular coordinate $\varphi$  describes rotations about the axis given by the vector $\vec{\mathsf{r}}$ that joins the two centers. $\vec{\mathsf{r}}$ establishes a natural length scale $\mathsf{r}=\vert \vec{\mathsf{r}}\vert$, so that the variables $\xi$ and $\eta$ are dimensionless.\\
Taking into account these definitions, an approximate electronic wave function  for a trilobite molecule \cite{eiles2019} can be written as  
\begin{equation}
\Upsilon^{(n)}_{\mathrm{t}}\Big(\xi, \eta; \vec{\mathsf{r}}\Big)=\frac{1}{\sqrt{\mathcal{N}_{\mathrm{t}}(\mathsf{r})}} \frac{u_{n,0} \big(\frac{\mathsf{r}}{2\mathrm{a}_0}(\xi+1)\big) u^{(1)}_{n,0} \big(\frac{\mathsf{r}}{2\mathrm{a}_0}(\eta+1)\big)-u^{(1)}_{n,0} \big(\frac{\mathsf{r}}{2\mathrm{a}_0}(\xi+1)\big) u_{n,0}\big(\frac{\mathsf{r}}{2\mathrm{a}_0}(\eta+1)\big)}{2 \pi \mathsf{r}(\xi-\eta)/\mathrm{a}_0},
\label{eq:to}
\end{equation}
whenever it has been generated from a Rydberg atom with an electron in a state with an effective principal quantum number $n$. In this equation $\mathrm{a}_0$ denotes the Bohr radius. This orbital involves  the  radial part of the hydrogenic wave function 
\begin{equation}
u_{nl}(r)=\sqrt{\left( \frac{2}{n} \right)^3 \frac{(n-l-1)!}{2n (n+l)!}} \left( \frac{2r}{n}\right)^l r e^{-r/n} L_{n-l-1}^{2l+1} \left( \frac{2r}{n}\right),
\label{eq:ufinal}
\end{equation}
and their $s$-th order derivative,$$u_{nl}^{(s)}(r) =\frac{d^su_{nl}(r)}{dr^s}.$$
The  electronic orbitals for a radial butterfly molecule are
\begin{eqnarray}
\Upsilon^{(n)}_{\mathrm{rb}}\Big(\xi, \eta; \vec{\mathsf{r}}\Big)&=&\frac{1}{\sqrt{\mathcal{N}_{\mathrm{rb}}(\mathsf{r})}} \Bigg[ \frac{\xi \eta-1}{2 \pi (\mathsf{r}/\mathrm{a}_0) ^2 (\xi-\eta)^3}\mathcal{F}\left(\frac{\mathsf{r}}{2\mathrm{a}_0}(\xi+1),\frac{\mathsf{r}}{2\mathrm{a}_0}(\eta+1) \right) \nonumber \\
&+& \frac{u_{n,0} \big(\frac{\mathsf{r}}{2\mathrm{a}_0}(\xi+1)\big) u^{(2)}_{n,0}\big(\frac{\mathsf{r}}{2\mathrm{a}_0}(\eta+1)\big)-u^{(2)}_{n,0} \big(\frac{\mathsf{r}}{2\mathrm{a}_0}(\xi+1)\big) u_{n,0}\big(\frac{\mathsf{r}}{2\mathrm{a}_0}(\eta+1)\big)}{4 \pi \mathsf{r} (\xi-\eta)/\mathrm{a}_0} \Bigg],\label{eq:abro}
\end{eqnarray}
while,  for a cosine angular butterfly molecule
\begin{equation}
\Upsilon^{(n)}_{\mathrm{ab}}\Big(\xi, \eta, \varphi; \vec{\mathsf{r}}\Big)=\frac{1}{\sqrt{\mathcal{N}_{\mathrm{ab}}(\mathsf{r})}} \frac{\sqrt{(\xi^2-1)(1-\eta^2)}}{2 \pi (\mathsf{r}/\mathrm{a}_0)^2 (\xi-\eta) ^3} \mathcal{F}\left(\frac{\mathsf{r}}{2\mathrm{a}_0}(\xi+1),\frac{\mathsf{r}}{2\mathrm{a}_0}(\eta+1) \right)  \cos \varphi.
\label{eq:abo}
\end{equation}
In Eqs.~(\ref{eq:abro}-\ref{eq:abo})
\begin{equation}
\begin{split}
\mathcal{F}(t_+,t_-)&= 2(t_--t_+) \,    u^{(1)}_{n,0}(t_+)  u^{(1)}_{n,0}(t_-) \\[0.15cm]
&- u_{n,0}(t_-)\left[ 2  u^{(1)}_{n,0}(t_+)- (t_+ - t_-) \,  u^{(2)}_{n,0}(t_+)  \right]+u_{n,0}(t_+)\left[ 2  u^{(1)}_{n,0}(t_-)+(t_+ - t_-) ,  u^{(2)}_{n,0}(t_-)  \right].
\end{split}\label{eq:ftptm}
\end{equation}
Both the trilobite and radial butterfly orbitals are invariant under rotations around the internuclear axis. Angular butterfly orbitals with well defined angular momentum along the internuclear axis depend on the azimuthal angle as $e^{\pm i\varphi}$ instead of $\cos\varphi$. For them the electronic probability density
is $\varphi$ independent. Expressions of the elementary LRRM orbitals could be considered the result of a mixed scheme since they entail spherical hydrogenic wave functions of spheroidal coordinates; however, the spheroidal nature of the problem becomes evident by observing the compact expressions
of the orbitals in terms of  products of functions of $\xi$ and of $\eta$. The peculiar structure of the trilobite and the butterfly Rubidium LRRM illustrated in Fig.~\ref{fig:regiones} for $n=33$, let observe that regions where the probability density is concentrated can be naturally enclosed by prolate spheroids whose foci lie on the neutral atom and the  Rydberg ionic core.  Enclosing spheres centered on that core
are also shown in Fig.~\ref{fig:regiones} for comparison. The radii of the spheres depends on the effective radial number $n$; the angular distribution is determined by the partial wave scattering factor (s- wave for trilobites and p-wave for butterfly molecules). In the shown examples, the spheroidal structure of trilobite is more evident than for butterfly orbitals. Nevertheless as $n$ increases the visibility of the spheroidal symmetry of butterfly electronic orbitals also increases. In all cases, the spheroidal nodal pattern of the wave functions can be interpreted as interference effects between four semiclassical elliptical trajectories focused on the Rydberg core and intersecting at both the neutral atom and at the observation point \cite{gay1988, granger2001}.

\section{LRRM electrostatic potential.}
The electrostatic potential $\phi(\vec{r})$ generated by  a charge distribution $\rho(\vec{r_s})$ in terms of
the unnormalized internal $V^{i}_{\ell m}(\vec r; \vec{\mathsf{r}})$ and external  $V^{e}_{\ell m}(\vec r; \vec{\mathsf{r}})$ spheroidal solid harmonics
is given by \cite{jansen2000}
\begin{eqnarray}
\phi(\vec{r})&=&\frac{2}{\mathsf{r}}\sum_{\ell=0}^{\infty} \sum_{m=-\ell}^{\ell} \int \rho(\vec{r_s}) (2\ell+1)(-1)^m \left[\frac{(\ell-m)!}{(\ell+m)!}\right]^2  V^{{i}*}_{\ell m}(\xi_<,\eta_s,\varphi_s;\vec{\mathsf{r}}) V^{e}_{\ell m}(\xi_>,\eta,\varphi; \vec{\mathsf{r}})d^3r_s,
\label{eq:pot1}\\
V^{i}_{\ell m}(\xi,\eta,\varphi; \vec{\mathsf{r}}) &=& \mathcal{P}_\ell^m(\xi)P_\ell^m(\eta)e^{im\varphi};\quad\quad V^{e}_{\ell m}(\xi,\eta,\varphi; \vec{\mathsf{r}})= Q_\ell^m(\xi)P_\ell^m(\eta)e^{im\varphi},
\end{eqnarray}
which involve the associated Legendre polynomials of the first kind, $P_\ell^m(x)$ ($\vert x\vert \le 1$) and $\mathcal{P}_\ell^m(x)$  ($\vert x\vert \ge 1$), and of the second kind $Q_\ell^m$. These functions are real when defined  by the Ferrer formulae
\begin{eqnarray}
P_\ell^m(x) = (1-x^2)^{m/2}\frac{d^m}{dx^m}P_\ell(x)&,& \vert x\vert \le 1,  0\le m \le \ell,\\
P_\ell(x) =\frac{1}{2^\ell \ell!}\frac{d^\ell}{dx^\ell}(x^2 -1)^\ell &,&\\
Q_\ell^m(x) = (-1)^m (x^2 -1)^{\vert m\vert/2}\frac{d^{m} }{dx^{m} }Q_\ell(x)&,&  x>  1, \vert m\vert \le \ell,\\
Q_\ell(x) =\frac{1}{2^\ell \ell!}\frac{d^\ell}{dx^\ell}\Big[(x^2 -1)^\ell\mathrm{ln}\frac{x+1}{x-1}\Big] &-&
\frac{1}{2}P_\ell(x)\mathrm{ln} \frac{x+1}{x-1},
\end{eqnarray}
for $m\ge 0$ and 
\begin{equation}
P_\ell^{-m}(x) = (-1)^m\frac{(\ell -m)!}{(\ell +m)!}P_\ell^{m}(x)
\end{equation}
otherwise.

The regular functions $\mathcal{P}_\ell^m(x)$ for $1\le x$ are obtained as
$$\mathcal{P}_\ell^m(x) = (x^2 -1)^{m/2}\frac{d^m}{dx^m}\mathcal{P}_\ell(x).$$
It can be shown \cite{Robin1957} that the Legendre functions of the second kind $Q_\ell^m(x)$ satisfy the inequality
\begin{equation}
\vert Q_\ell^m(x) \vert< \frac{2^{\ell +1} (\ell -m)!}{(2\ell +1)!!}\Big(\frac{1}{x+\sqrt{x^2-1}}\Big)^{\ell +1}\Big[\frac{1}{2}\Big(\Big( \sqrt{\frac{x+1}{x-1}}\Big)^m+\Big( \sqrt{\frac{x-1}{x+1}}\Big)^m\Big)\Big(\frac{x+\sqrt{x^2-1}}{2\sqrt{x^2 -1}}\Big)\Big].
\end{equation}
So that, for $x>>1$ these functions decay faster than $x^{-(\ell +1)}$.

In these expressions, the dependence on the vector $\vec{\mathsf{r}}$ that joins the spheroidal foci is implicit in the definition of the coordinates $\vec r:\{\xi,\eta,\varphi\}$.

Spheroidal solid harmonics are analogous to the spherical solid harmonics
\begin{equation}
U^{i}_{\ell m}(\vec r) =  r^{\ell} P_\ell^m(\cos\theta)e^{im\varphi};\quad\quad U^{e}_{\ell m}(\vec r)= r^{-(\ell+1)} P_\ell^m(\cos\theta)e^{im\varphi},
\end{equation}
defined in spherical coordinates $\vec r: \{r,\theta,\varphi\}$ in standard notation.
In this coordinates there is no natural length scale.
Some properties of the spheroidal solid harmonics are summarized in the Appendix, in particular their connection to spherical solid harmonics is given; for $\ell=0,1$ internal spherical and spheroidal harmonics coincide.

\begin{figure}[H]
\centering
\begin{tabular}{@{}c @{}c @{}c}
\includegraphics[width=.33\textwidth]{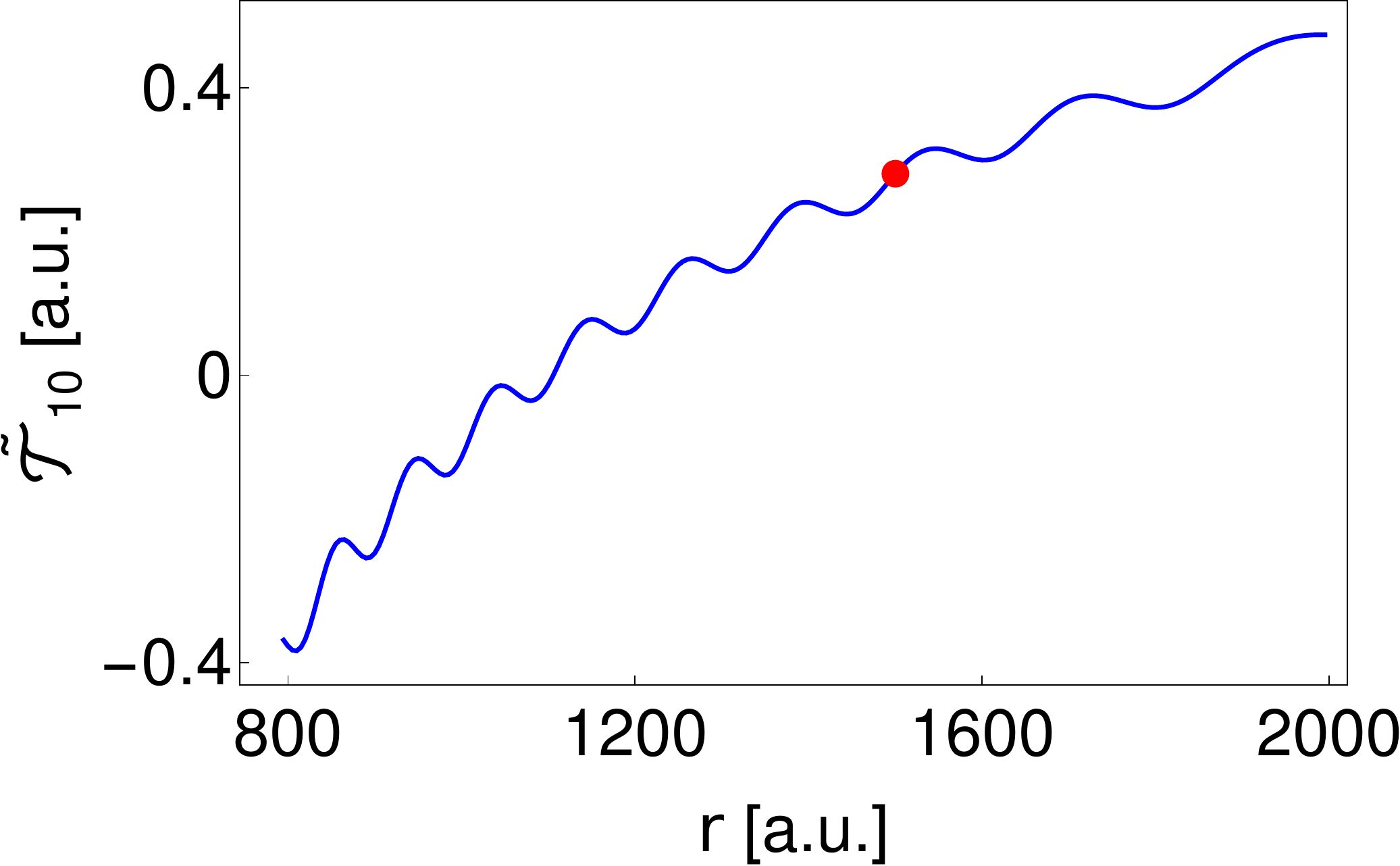}&
\includegraphics[width=.33\textwidth]{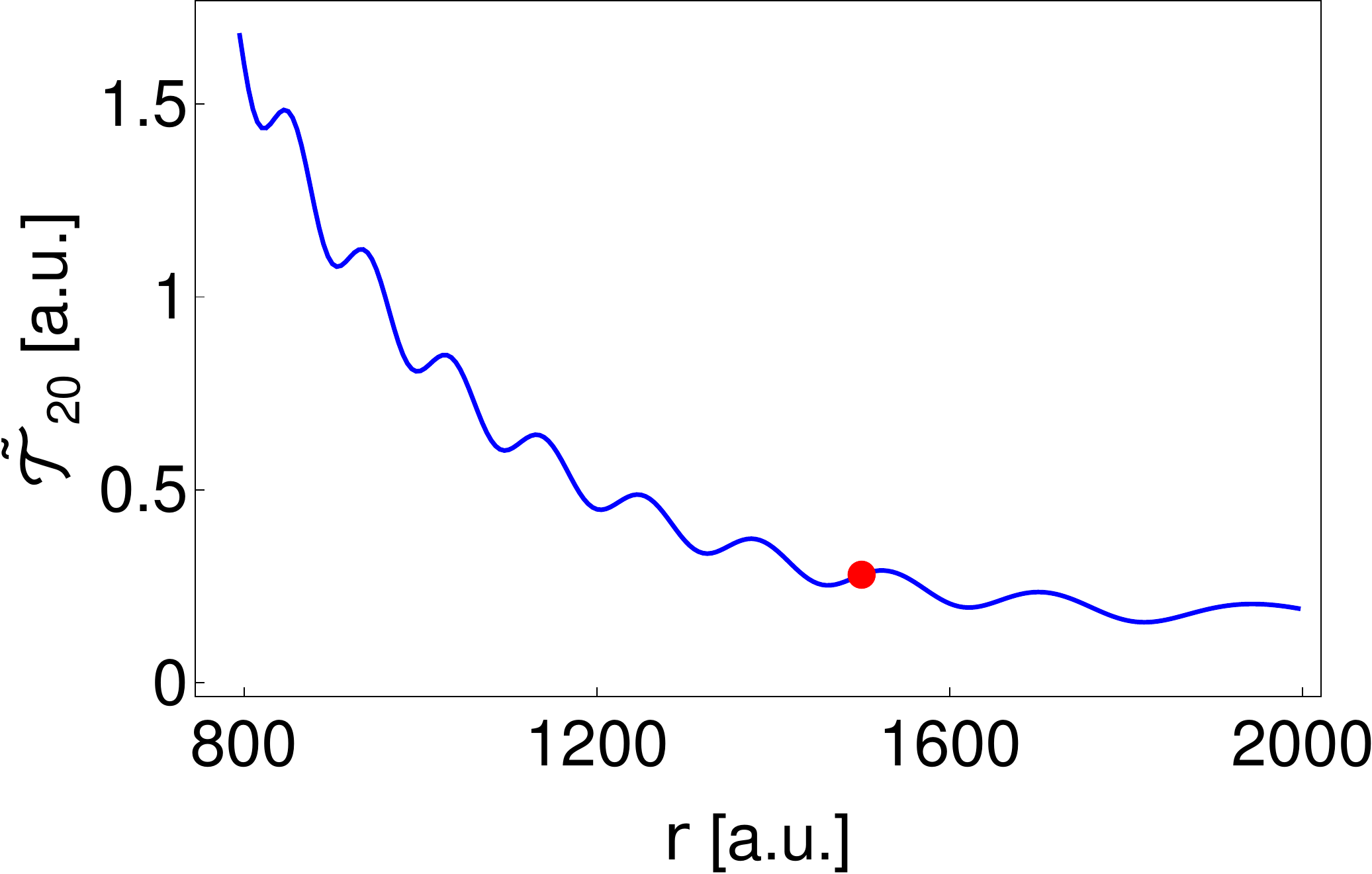}&
\includegraphics[width=.33\textwidth]{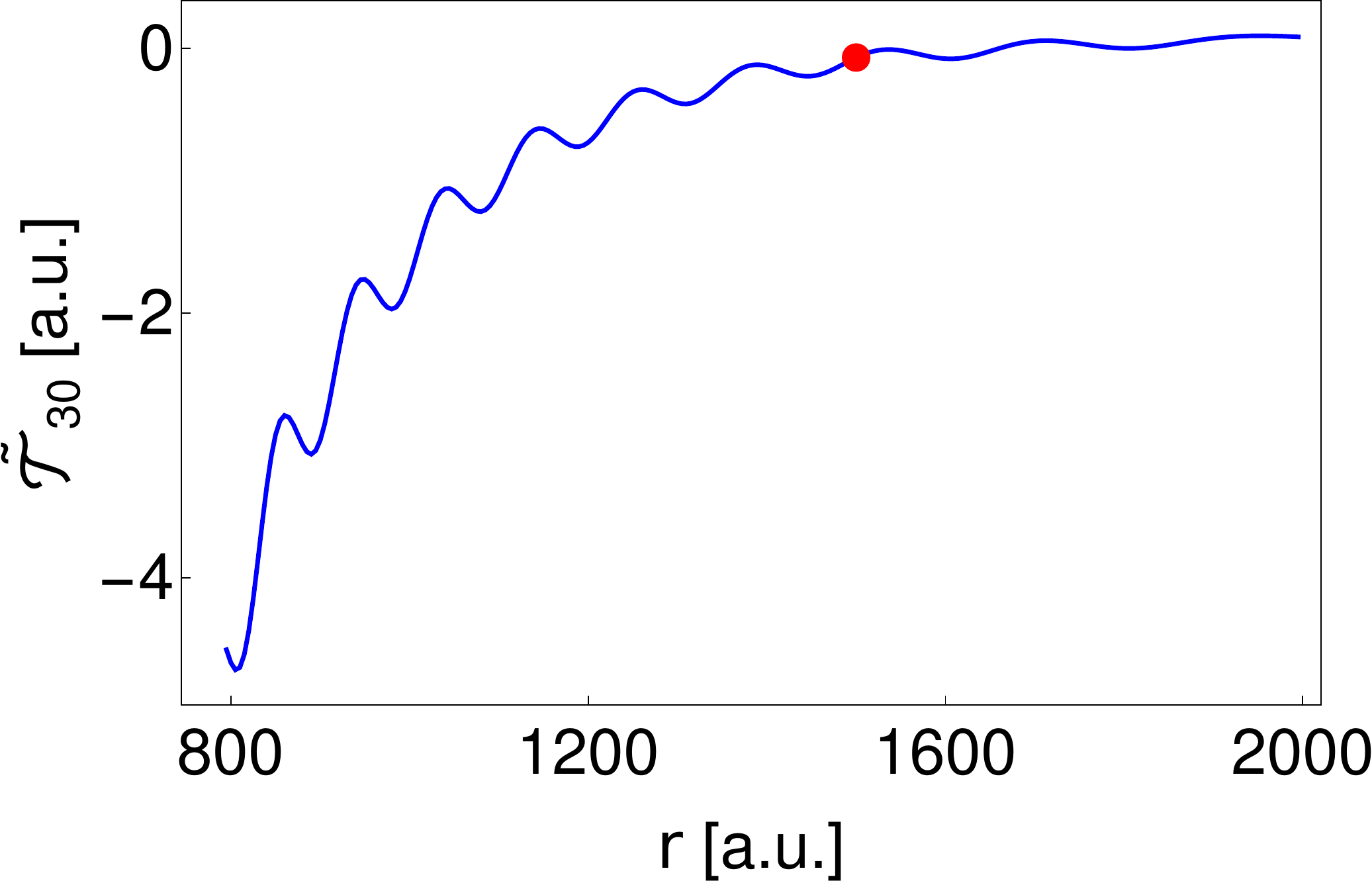}\\
(Ia)&(Ib)&(Ic)\\
\includegraphics[width=.33\textwidth]{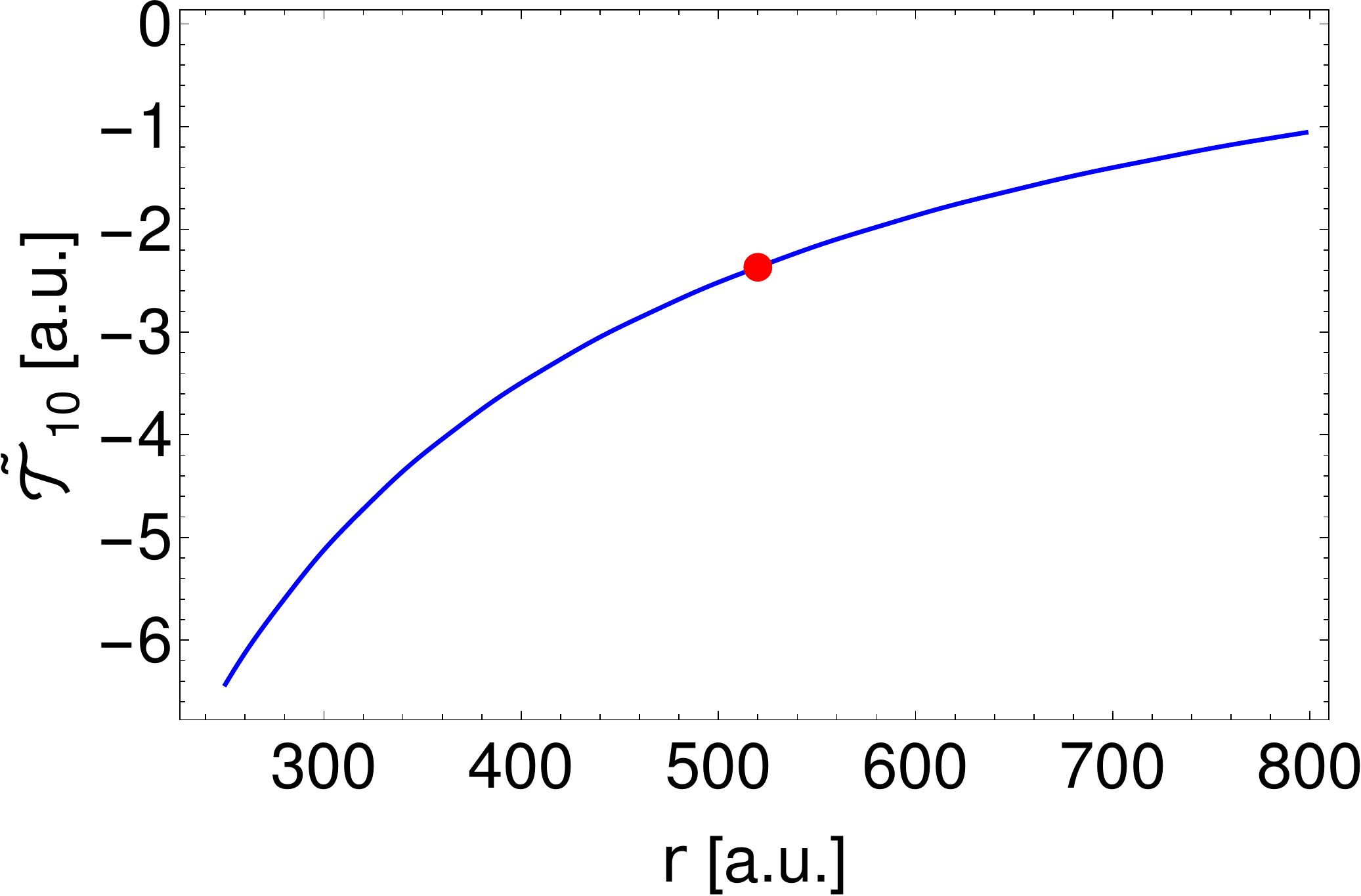}&
\includegraphics[width=.33\textwidth]{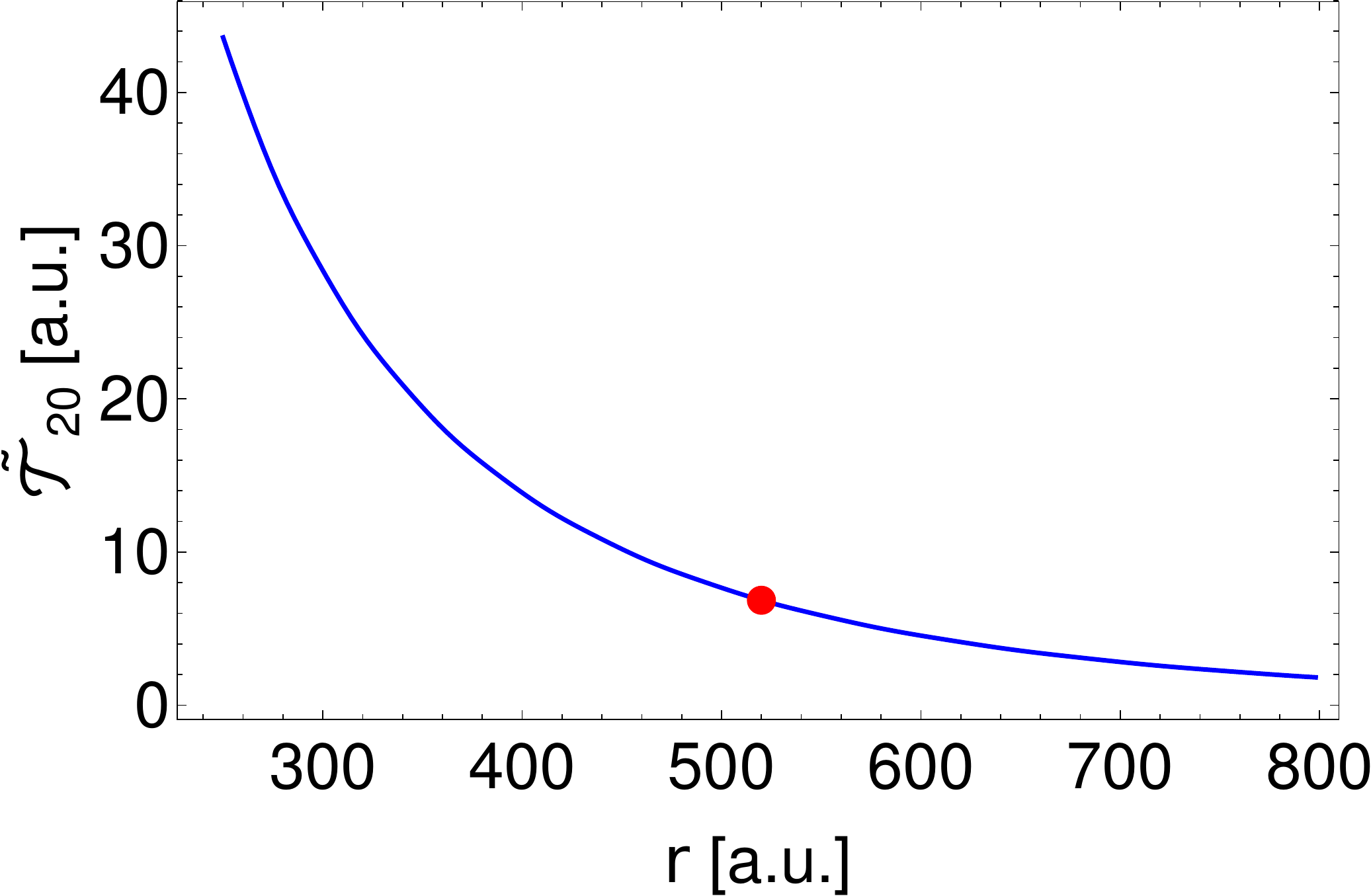}&
\includegraphics[width=.33\textwidth]{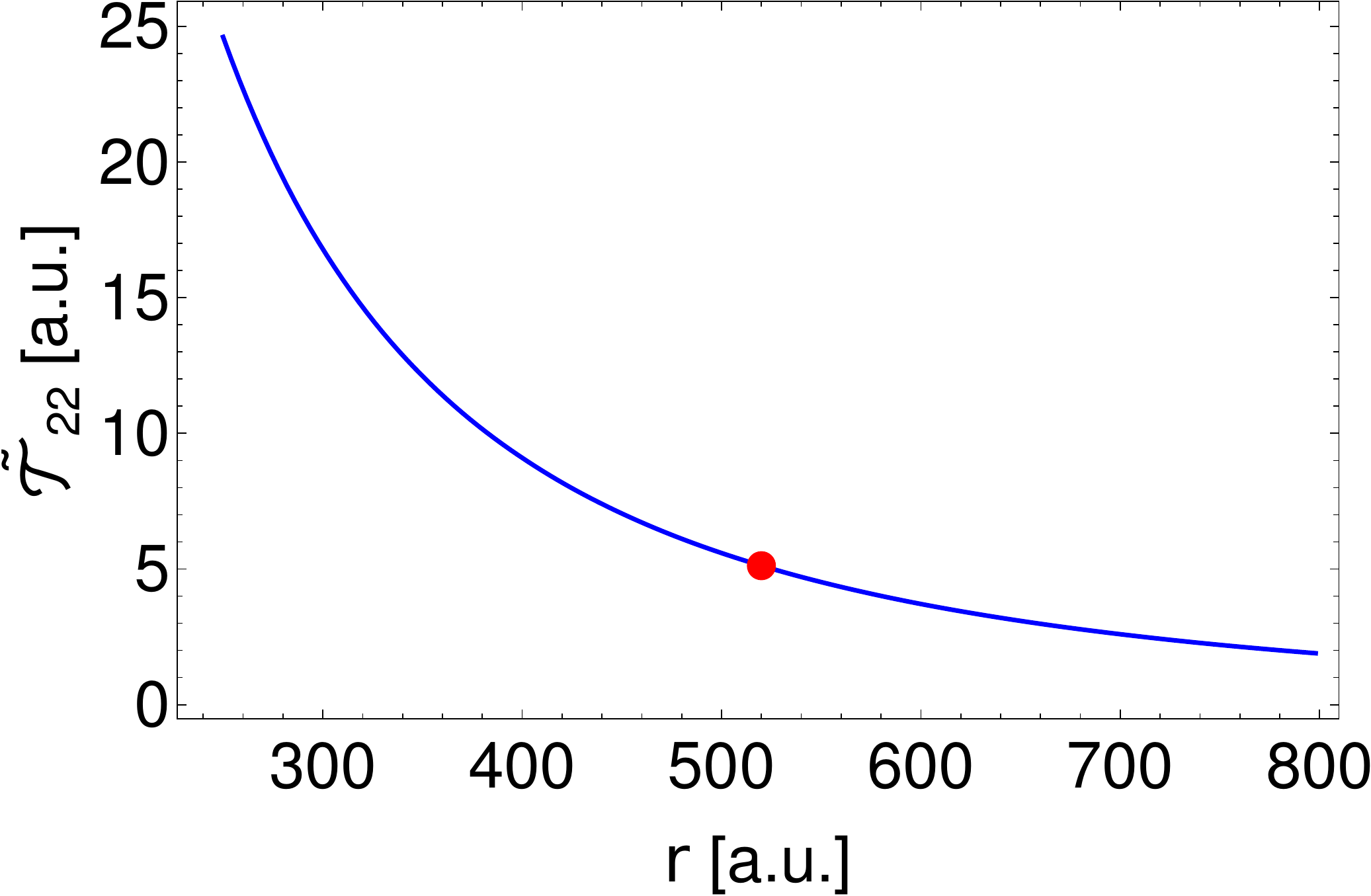}\\
(IIa)&(IIb)&(IIc)
\end{tabular}
\caption{ Spheroidal internal dimensionless multipole moments of Rb LRRM with $n =33$ as a function of the internuclear distance ${\mathsf{r}}$ for: (I) a trilobite orbital  (Ia) $\widetilde{\mathcal{T}}_{1,0}$, (Ib) $\widetilde{\mathcal{T}}_{2,0}$ and (Ic) $\widetilde{\mathcal{T}}_{3,0}$ ; (II) a cosine butterfly (IIa) $\widetilde{\mathcal{T}}_{1,0}$, (IIb) $\widetilde{\mathcal{T}}_{2,0}$ and (IIc) $\widetilde{\mathcal{T}}_{2,2}$.  Length is measured  in atomic units and the red mark corresponds to the stable configuration $\mathsf{r} =1500$a$_0$ for the trilobite molecule and
$\mathsf{r} =520$a$_0$ for the cosine butterfly molecule.
}\label{fig:fig1}
\end{figure}

For localized charge distributions, the spheroidal internal multipole moments
\begin{eqnarray}
\mathcal{T}^{i}_{\ell m}(\vec{\mathsf{r}}) &=&\Big(\frac{\mathsf{r}}{2}\Big)^\ell \widetilde{\mathcal{T}}^{i}_{\ell m}(\vec{\mathsf{r}})\\
                         &=&\Big(\frac{\mathsf{r}}{2}\Big)^\ell \frac{(\ell -m)!}{(2\ell -1)!!}
\sqrt{\frac{(\ell -m)!}{(\ell +m)!}}\int_{\mathbb{R}^3}
\rho(\vec{r}_s) V_{\ell m}^i(\vec{r}_s;\vec{\mathsf{r}})d^3r_s \label{eq:multipole}
\end{eqnarray}
provide an electrostatic characterization of the source. 
At any point outside the smallest prolate spheroid $\xi=\xi_0$ that encloses the charge distribution, 
\begin{eqnarray}
\phi(\vec{r})&=&
\sum_{\ell=0}^{\infty} \sum_{m=-\ell}^{\ell} \mathcal{I}_{\ell m}^*(\vec{r};\vec{\mathsf{r}}) \mathcal{T}^{i}_{\ell m}(\vec{\mathsf{r}})
\label{eq:pots}\\
&=&\frac{2}{\mathsf{r}}
\sum_{\ell=0}^{\infty} \sum_{m=-\ell}^{\ell} \widetilde{\mathcal{I}}_{\ell m}^*(\vec{r};\vec{\mathsf{r}}) \widetilde{\mathcal{T}}^{i}_{\ell m}(\vec{\mathsf{r}})\\
\mathcal{I}_{\ell m}(\vec{r};\vec{\mathsf{r}})& =& \Big(\frac{2}{\mathsf{r}}\Big)^{\ell+1}\widetilde{\mathcal{I}}_{\ell m}(\vec{r};\vec{\mathsf{r}})\\& =&
\Big(\frac{2}{\mathsf{r}}\Big)^{\ell+1} \frac{(2\ell+1)!!}{(\ell+m)!} \sqrt{\frac{(\ell-m)!}{(\ell+m)!}} V_{\ell m}^e(\vec r;\vec{\mathsf{r}}).
\end{eqnarray}
The spheroidal multipole moments associated to a LRRM can be calculated in terms of the expectation value of $V^i_{\ell m}(\vec r;\vec{\mathsf{r}})$ for the corresponding electronic orbital. We illustrate in Fig.~\ref{fig:fig1} the first three non-trivial moments as a function of the nuclear separation $\mathsf{r}$ for the electron orbital of a trilobite and a cosine butterfly molecule. The origin of coordinates was taken at the core of the Rydberg atom. A stable configuration for Rb $n=33$ trilobite molecule corresponds to
${\mathsf{r}} = 1500$a$_0$ and for a cosine butterfly ${\mathsf{r}} = 520$a$_0$, which are pointed out using a red mark in the Figure. Notice  that for the trilobite the corresponding value of the dimensionless multipole moments $\widetilde{\mathcal{T}}_{1,0}=0.2806$, $\widetilde{\mathcal{T}}_{2,0}=0.2798$, and $\widetilde{\mathcal{T}}_{3,0}=-0.0710$; that is, the dimensionless dipole and quadrupole moments have a similar value which prevents neglecting any of them in a given physical scenario before making a careful analysis. It must also be kept in mind that for $\ell=0,1$ internal spherical and spheroidal moments $\mathcal{T}^{i}_{\ell m}$ coincide. In the case of the  cosine butterfly molecule the structure of the momenta is found to be monotonic as a function of $\mathsf{r}$, and at the internuclear separation that yields stability to the molecule $\widetilde{\mathcal{T}}_{1,0}=- 2.369$, $\widetilde{\mathcal{T}}_{2,0}=6.848$, and $\widetilde{\mathcal{T}}_{2,2}=5.122$. That is, the dimensionless quadrupole moment is even greater than the dipole one. This means that the latter cannot be neglected in a reliable description of the molecule electrostatic potential and the intermolecular interactions.  

The electrostatic potential $\phi_{LRRM}(\vec{r})$ generated by a LRRM, arises from the charge distribution 
\begin{equation*}
\rho(\vec{r}; \vec{\mathsf{r}})=\delta(\vec{r})-\left| \Upsilon_\beta(\vec{r}; \vec{\mathsf{r}}) \right|^2.
\end{equation*}
that includes both the Rydberg core, which we are modeling as a positive unit point charge located at the origin, and the Rydberg electron in an orbital $\Upsilon_\beta^{(n)}$ with an associated charge density $\rho^{(n)}_\beta$. As a consequence,
\begin{equation}
\phi_{LRRM}(\vec{r})=\frac{1}{r}-\int  \frac{\rho^{(n)}_\beta(\vec r_s;\vec{\mathsf{r}})}{|\vec{r}-\vec{r_s}|}  d^3r_s.
\label{eq:exa}
\end{equation}
 Notice that in a variational or self consistent description of a LRRM in the presence of an additional electric field, it would be adequate to take $\Upsilon_{\beta}$  as a linear superposition of the elementary orbitals $\Upsilon_{\beta}^{(n)}$ described by Eqs.~(\ref{eq:to}-\ref{eq:ftptm}). 

The relevance of multipole expansions relays on the possibility of truncating the series to a finite order that guarantees a reliable potential evaluation to a given precision in a previously chosen spatial region.
In the next paragraphs, we  compare $\phi(\vec{r})$ evaluated using the exact expression Eq.~(\ref{eq:exa}) to that resulting from
the multipole spheroidal and spherical expansions up to dipole terms; although the  monopole and internal dipole moments for spherical and prolate spheroidal coordinates coincide, their corresponding electrostatic potentials differ since
\begin{eqnarray}
V_{00}^e(\vec r;\vec{\mathsf{r}}) =\frac{1}{2}\mathrm{ln}\Big(\frac{\xi +1}{\xi -1}\Big) &,& 
V_{10}^e(\vec r;\vec{\mathsf{r}}) =\eta\Big( -1 + \frac{\xi}{2}\mathrm{ln}\Big(\frac{\xi +1}{\xi -1}\Big)\Big),\\
U_{00}^e(\vec r) =\frac{1}{r}&,& U_{10}^e(\vec r) =\frac{\cos\theta}{r^2}.
\end{eqnarray}
The symmetry under rotations around the internuclear axis of trilobite, radial and $m=\pm 1$ angular butterfly electronic orbitals, allows to visualize the potential general behavior by calculating it at any plane that contains the $z$ axis. As a figure of merit,  the relative error given by 
\begin{equation}
\left| \frac{\phi_{LRRM}(\vec{r})-\phi^{(\ell)}(\vec{r})}{\phi_{LRRM}(\vec{r})} \right|,\label{eq:rel}
\end{equation}
for expansions that consider up to an $\ell$ term, is also shown.
Taking into account the structure of multipole expansions, three different space regions are considered. The near region which here is taken as that which extends from the Rydberg atom nucleus to a couple of times the internuclear separation ($\sim 3000 \, \mathrm{a}_0$ for trilobite Rb molecules with $n=33$). There, details of the structure  of the electronic orbital are more evident. The second region is the far region, here taken at about $12 \cdot{\mathsf{r}}$ ($\sim 18000 \, \mathrm{a}_0$).  This region is particularly important for the description of electrostatic interactions in dilute gases. Finally we study an intermediate region ($\sim 7000 \, \mathrm{a}_0$ from the Rydberg core). \\

\begin{figure}[h!]
\centering

\begin{tabular}{@{}c  @{}c  @{}c  @{}c @{}c @{}c}
\includegraphics[width=.06\textwidth]{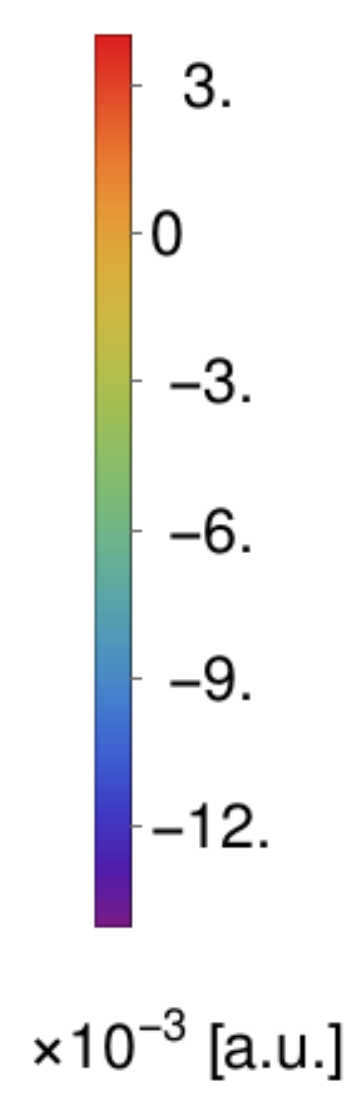}&
\includegraphics[width=.21\textwidth]{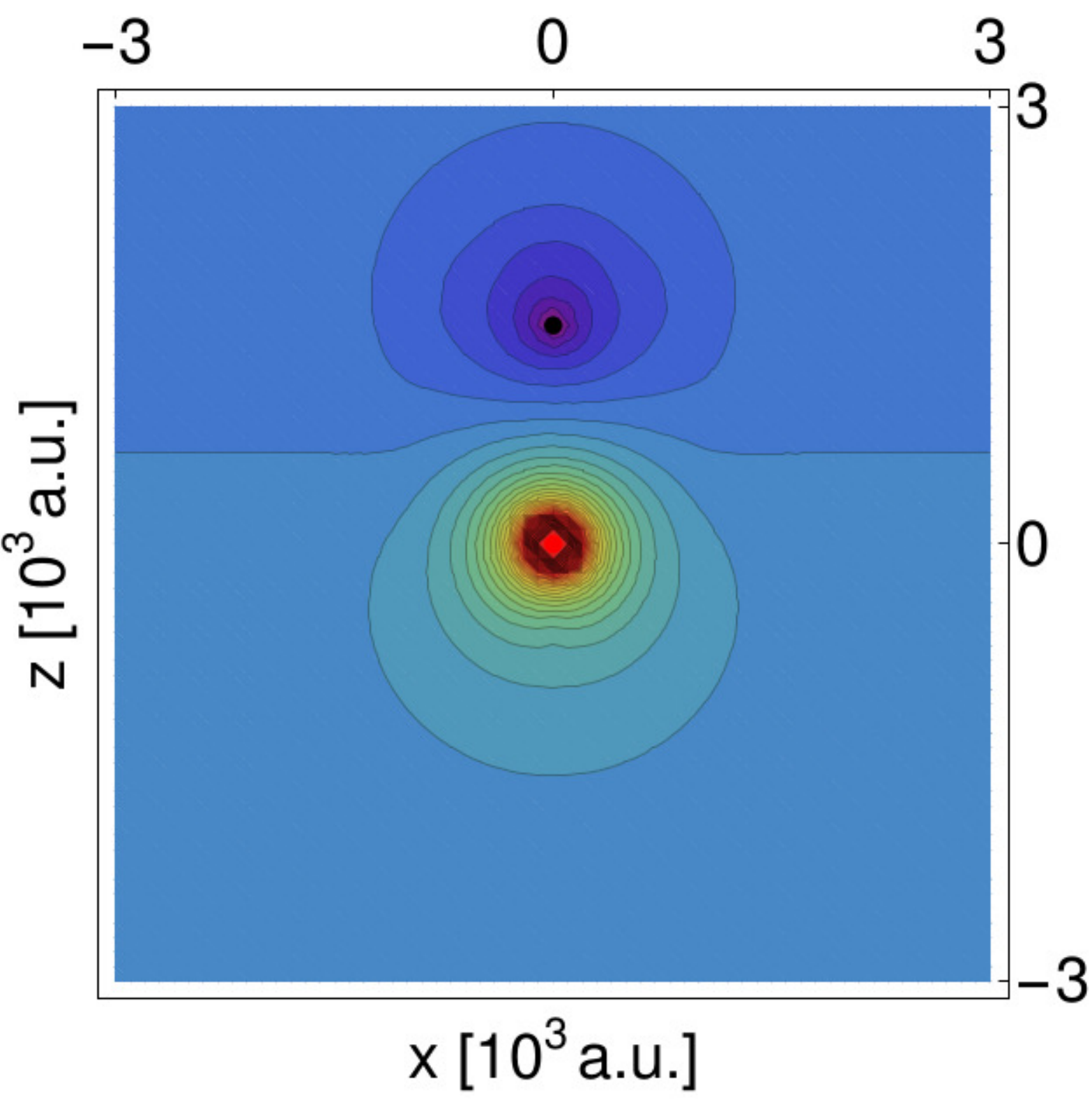}&
\includegraphics[width=.21\textwidth]{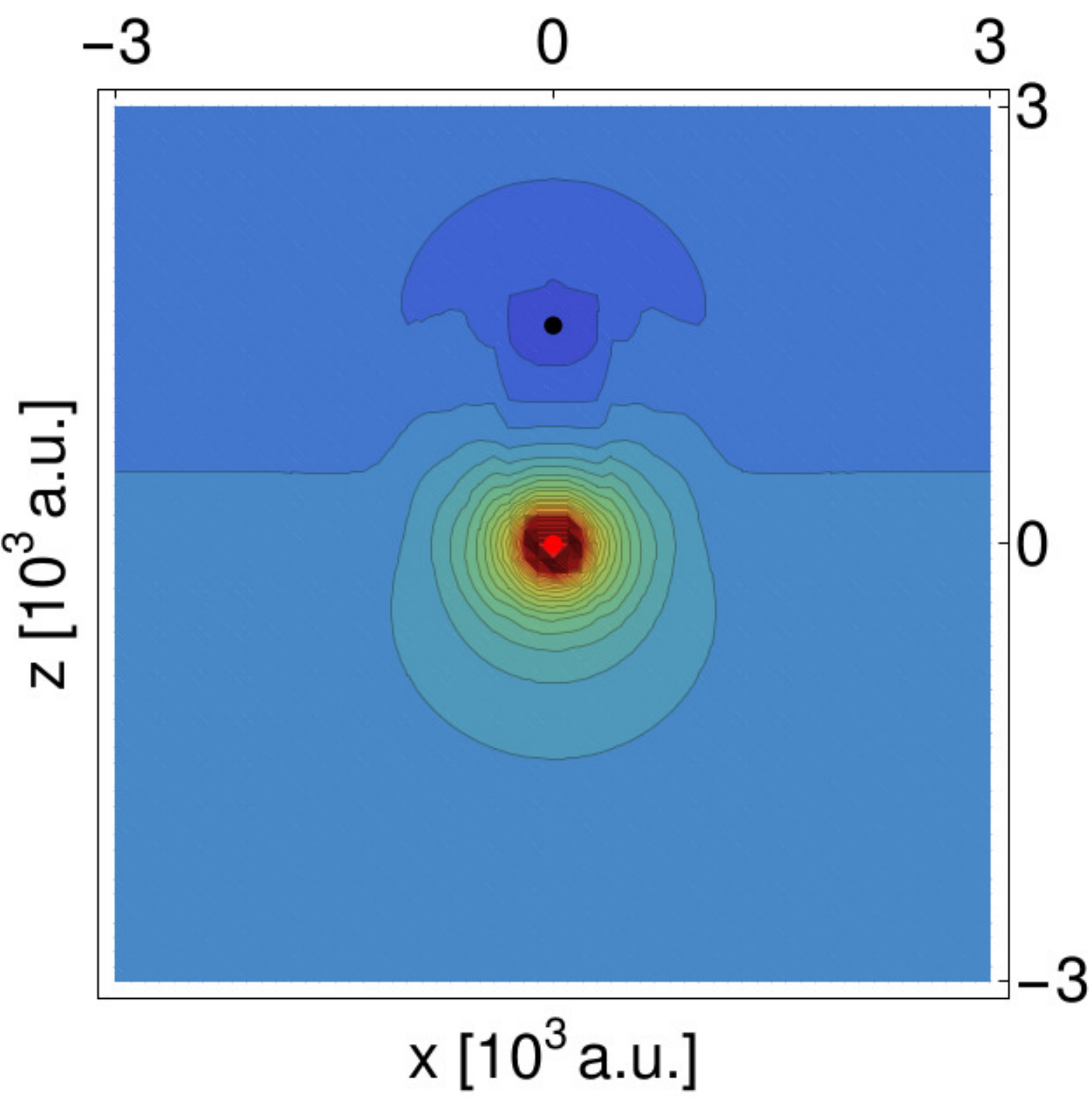}&
\includegraphics[width=.21\textwidth]{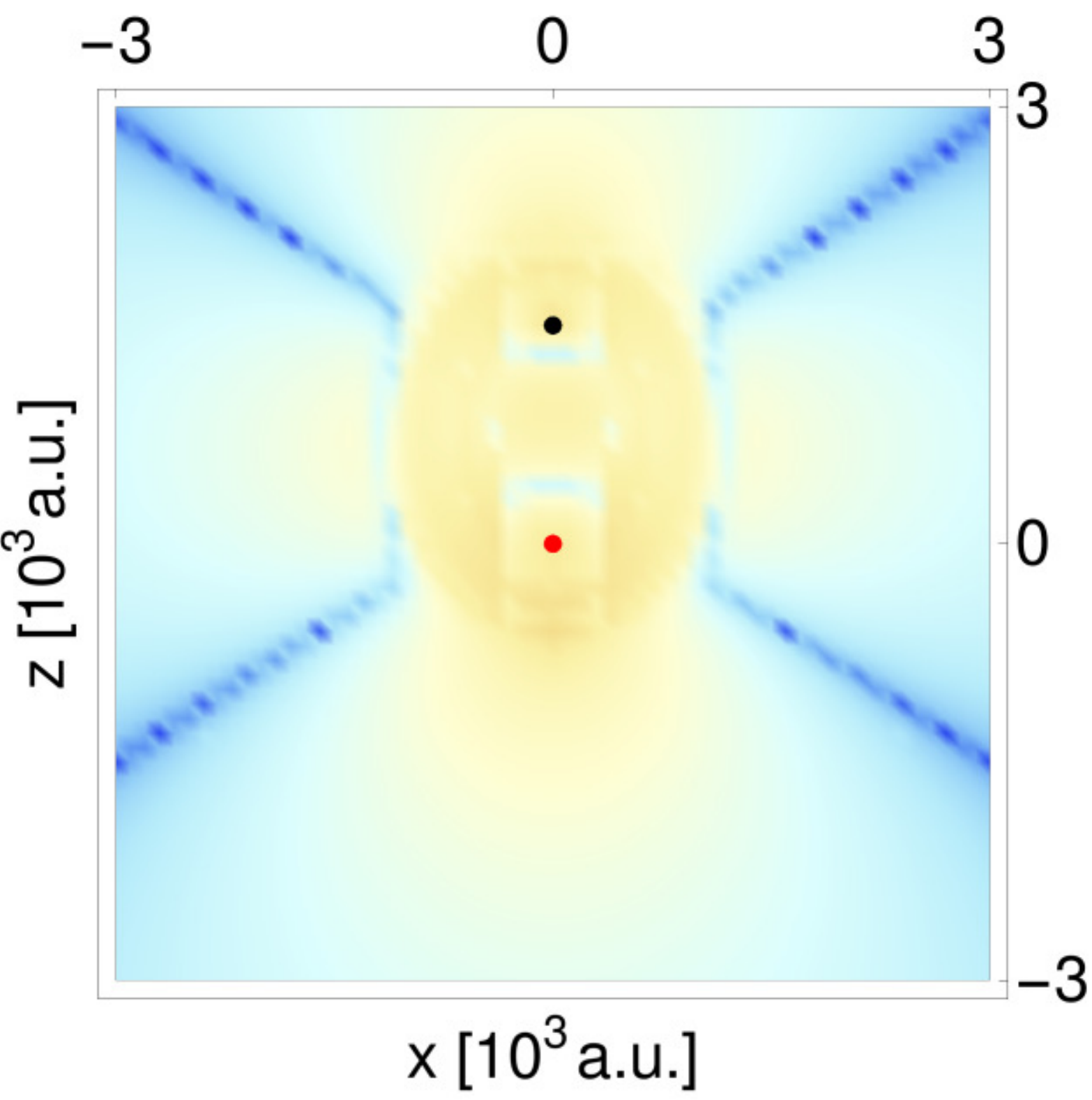}&
\includegraphics[width=.21\textwidth]{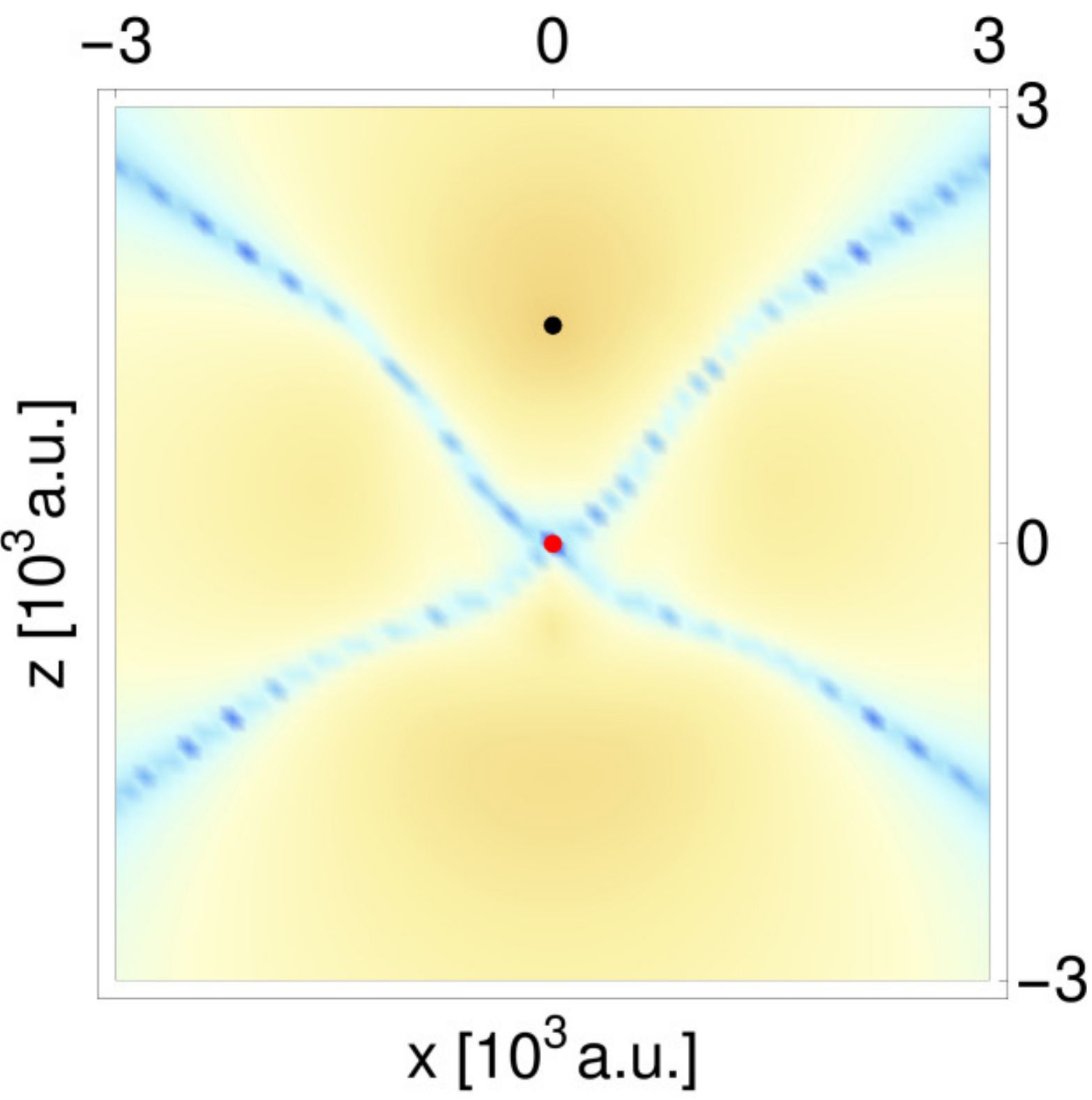}&
\includegraphics[width=.032\textwidth]{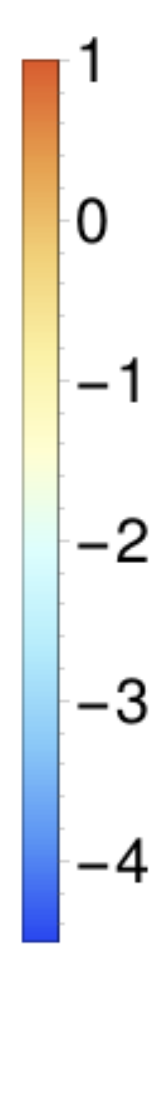}\\
& (a) &(b) &(c) &(d) & \\
\includegraphics[width=.06\textwidth]{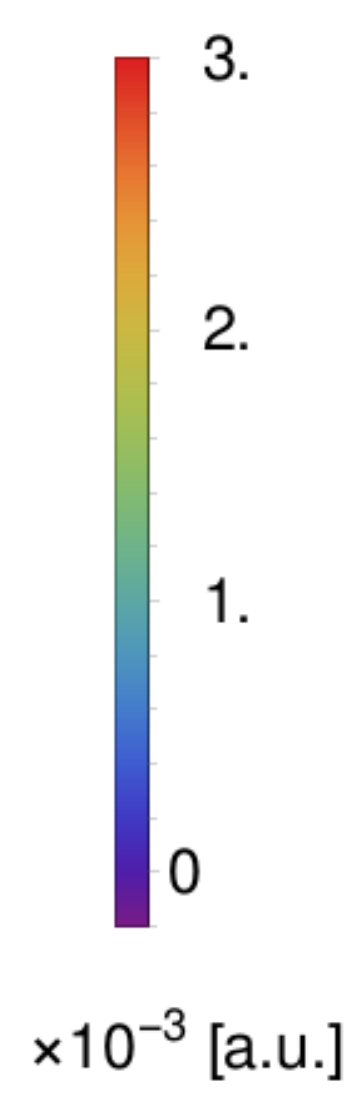}&
\includegraphics[width=.21\textwidth]{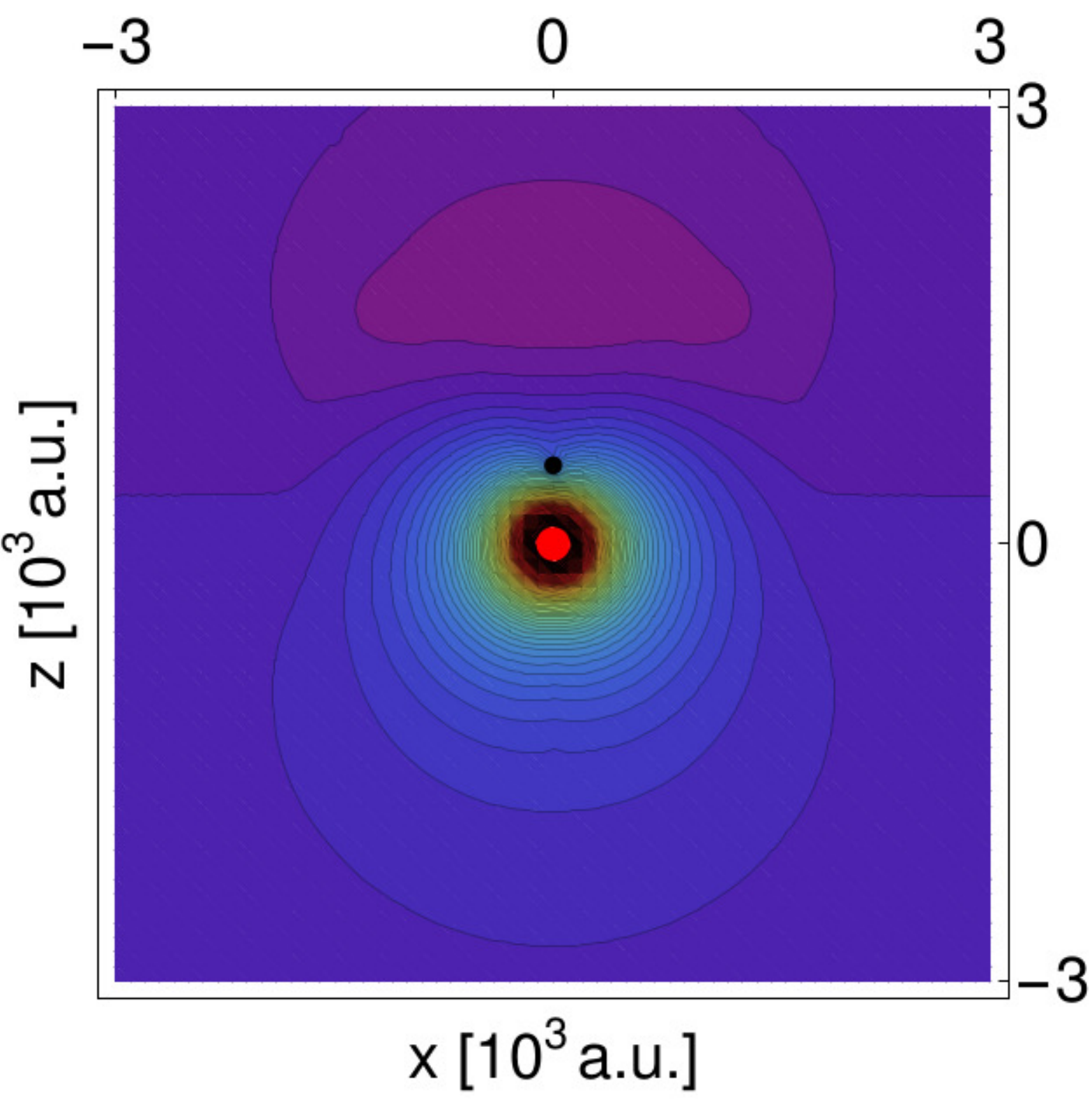}&
\includegraphics[width=.21\textwidth]{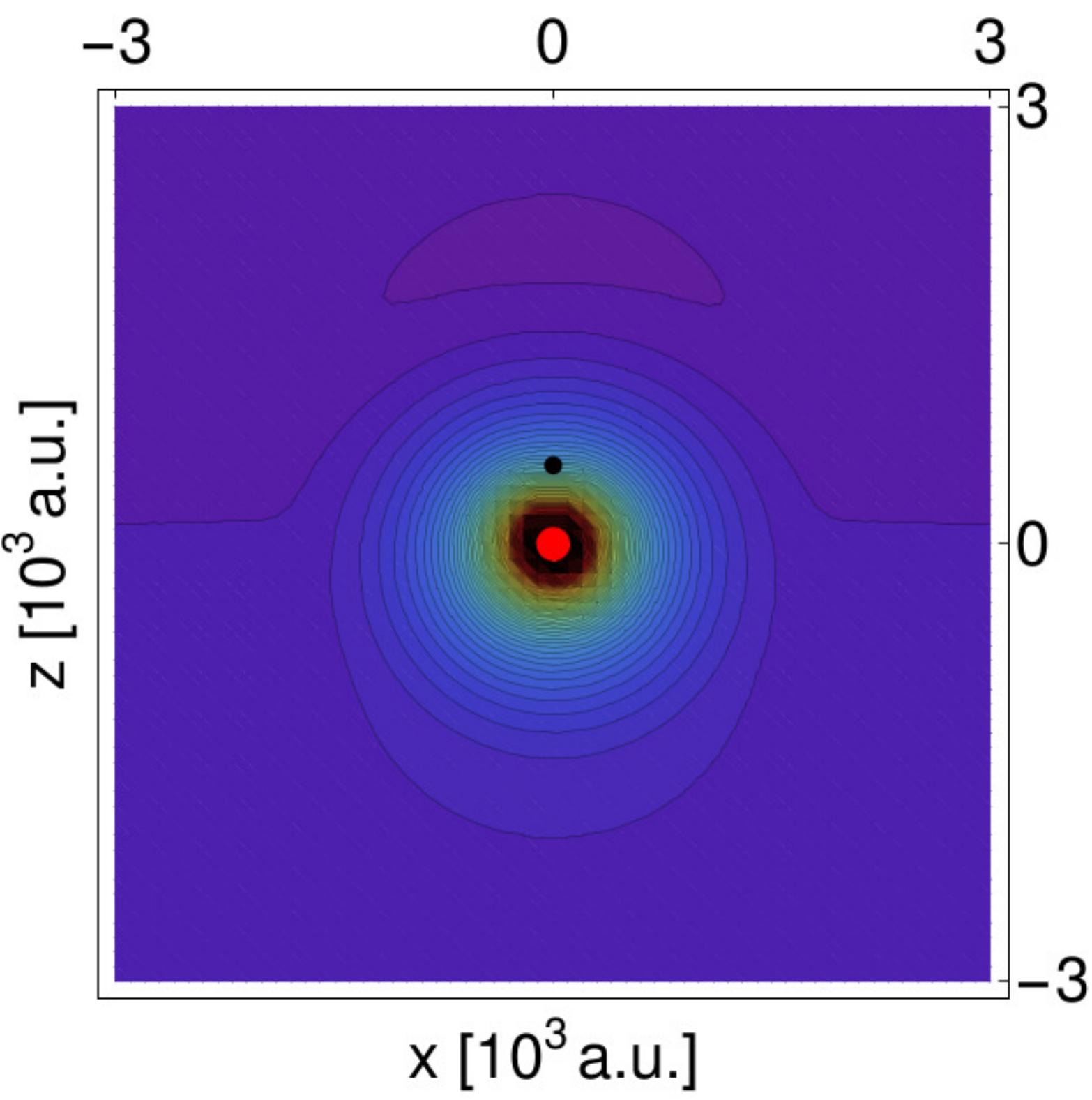}&
\includegraphics[width=.21\textwidth]{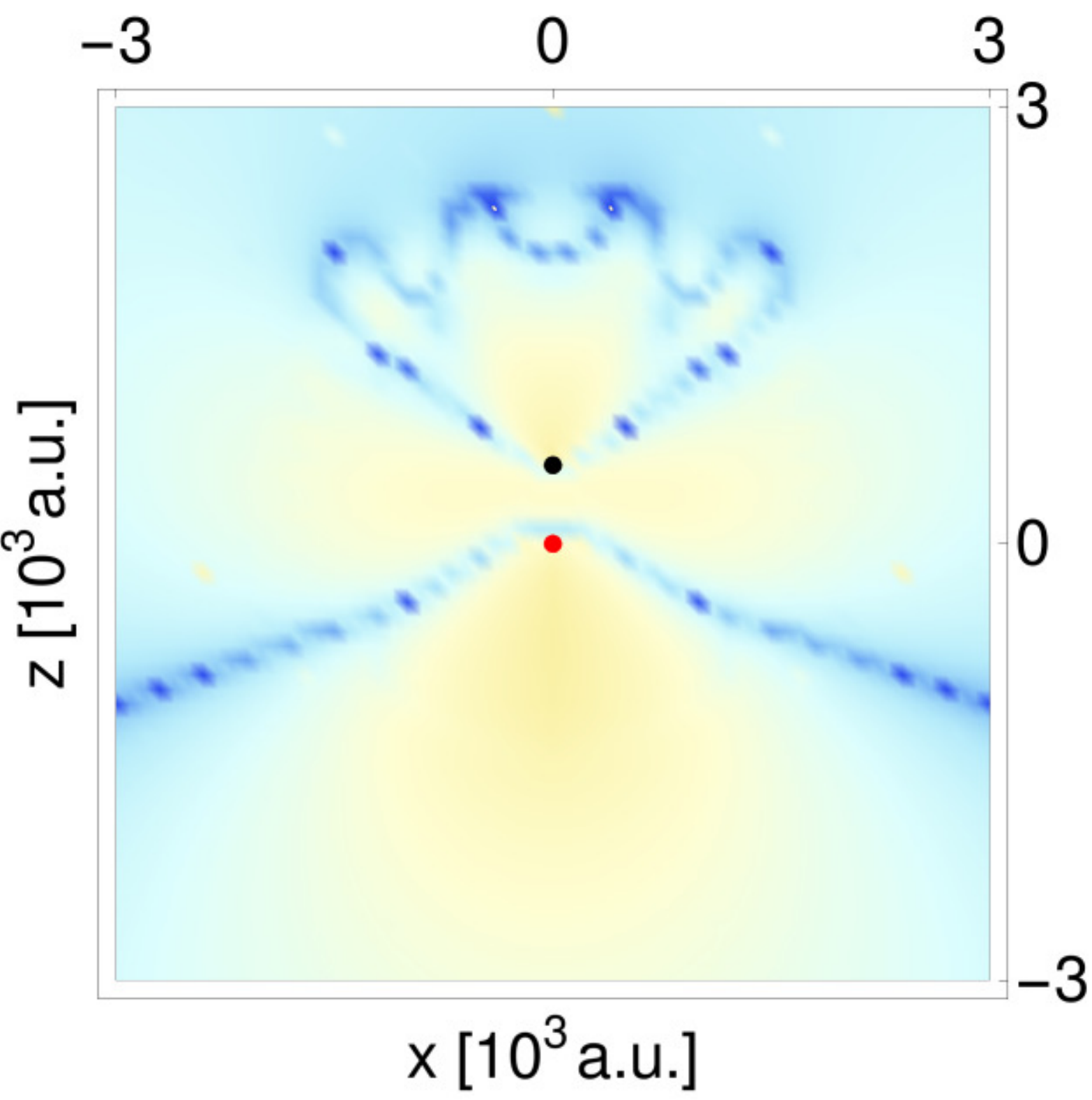}&
\includegraphics[width=.21\textwidth]{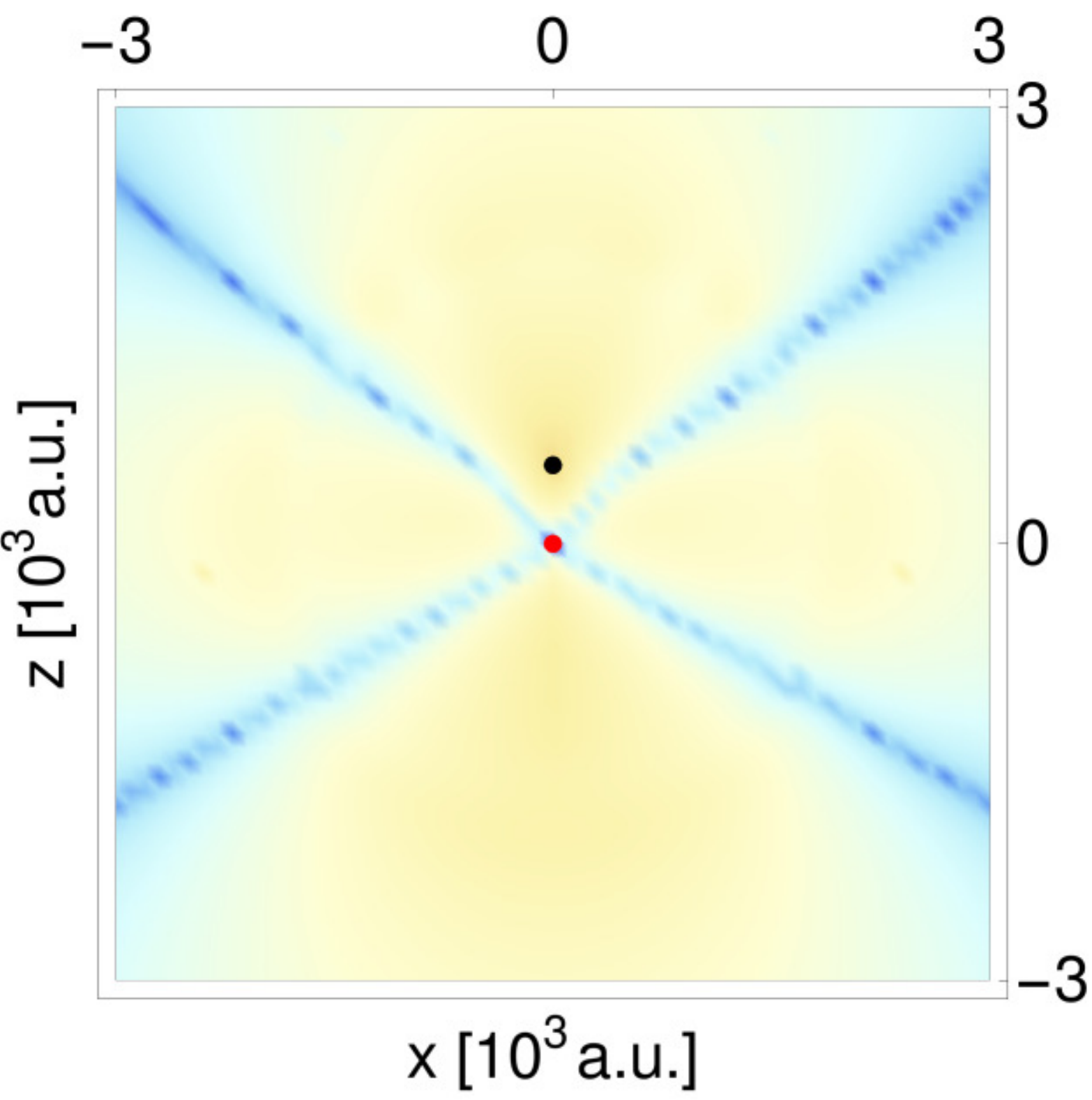}&
\includegraphics[width=.03\textwidth]{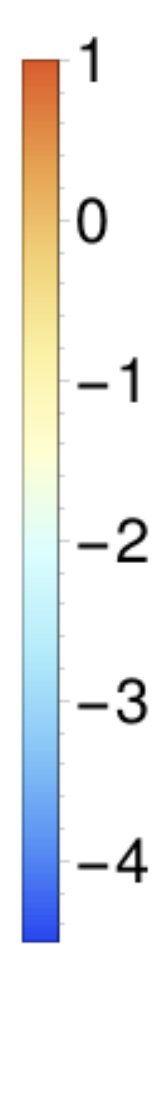}\\
&(e) & (f) &(g) &(h) & \\
\includegraphics[width=.06\textwidth]{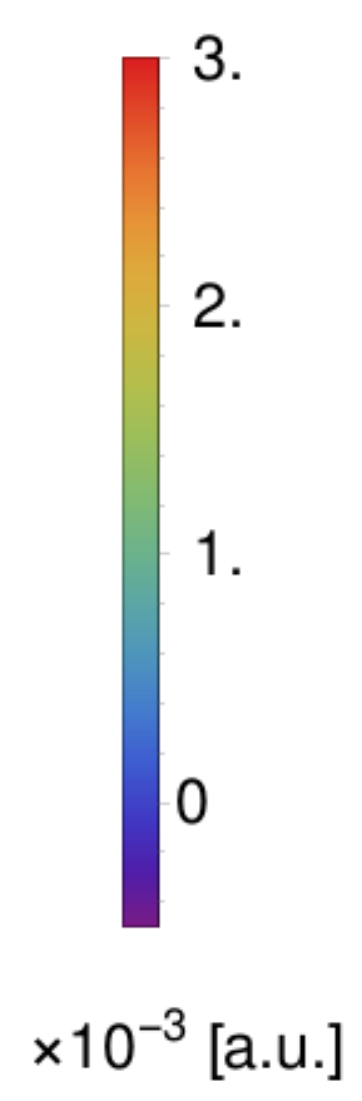}&
\includegraphics[width=.21\textwidth]{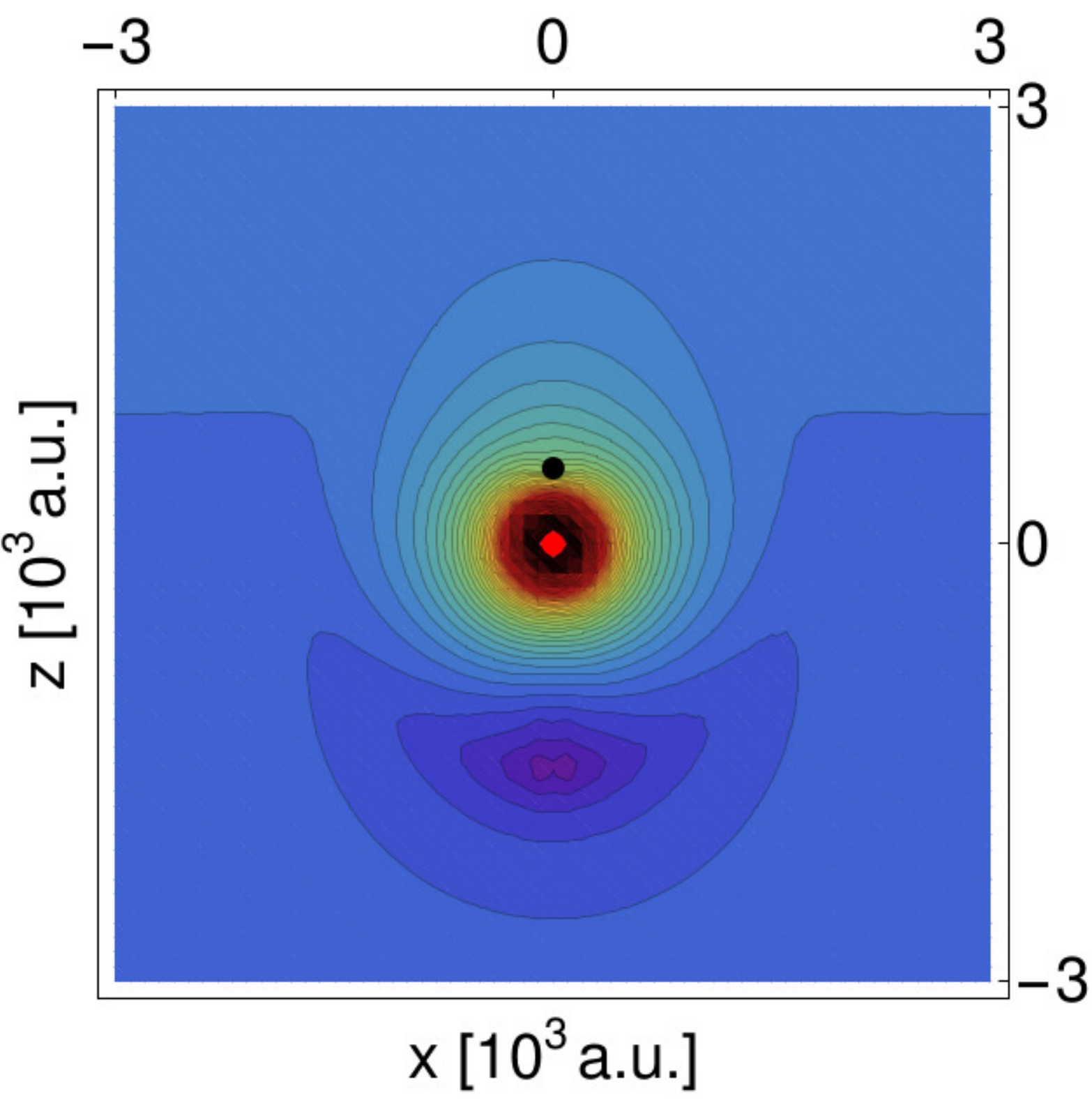}&
\includegraphics[width=.21\textwidth]{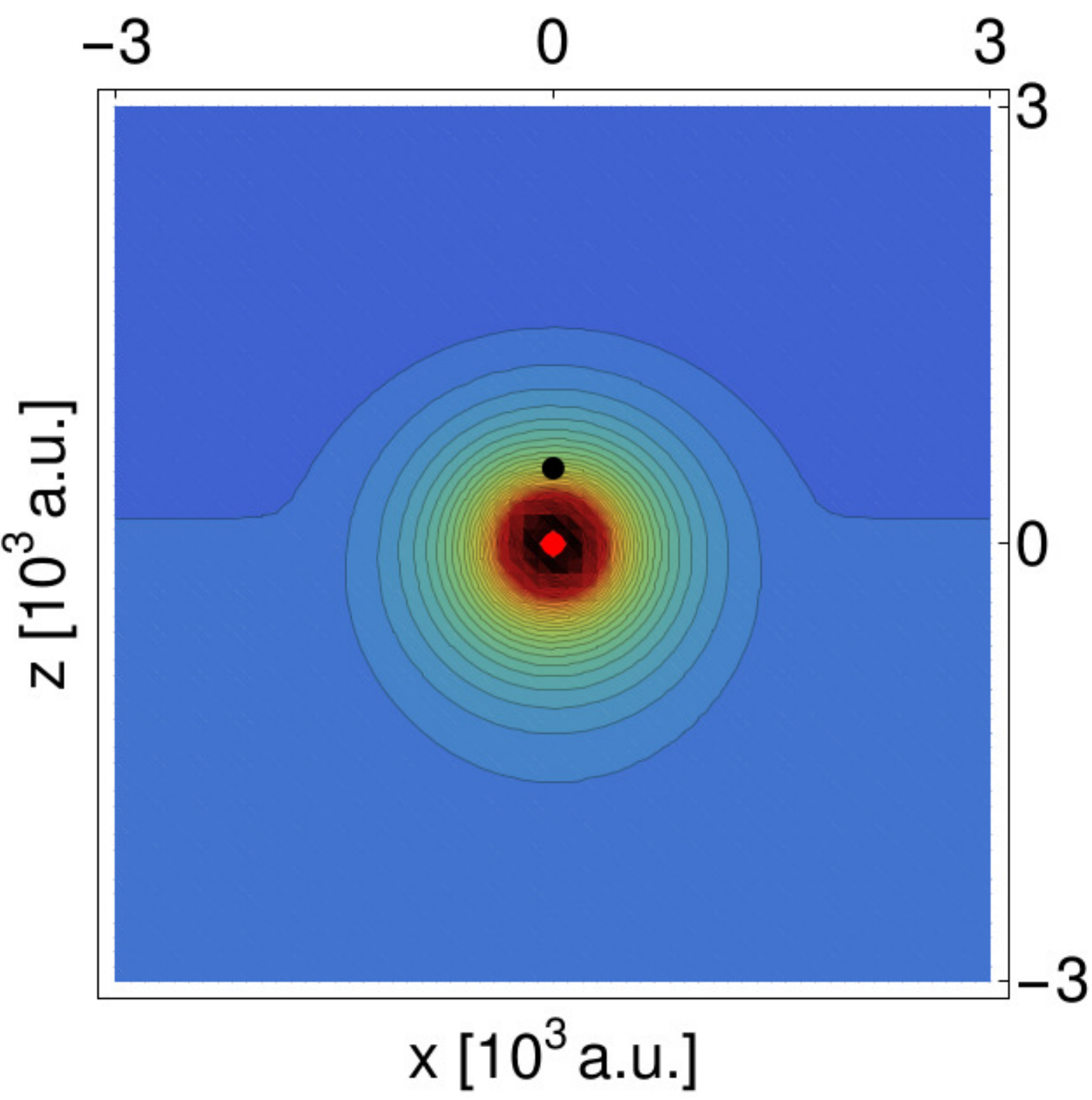}&
\includegraphics[width=.21\textwidth]{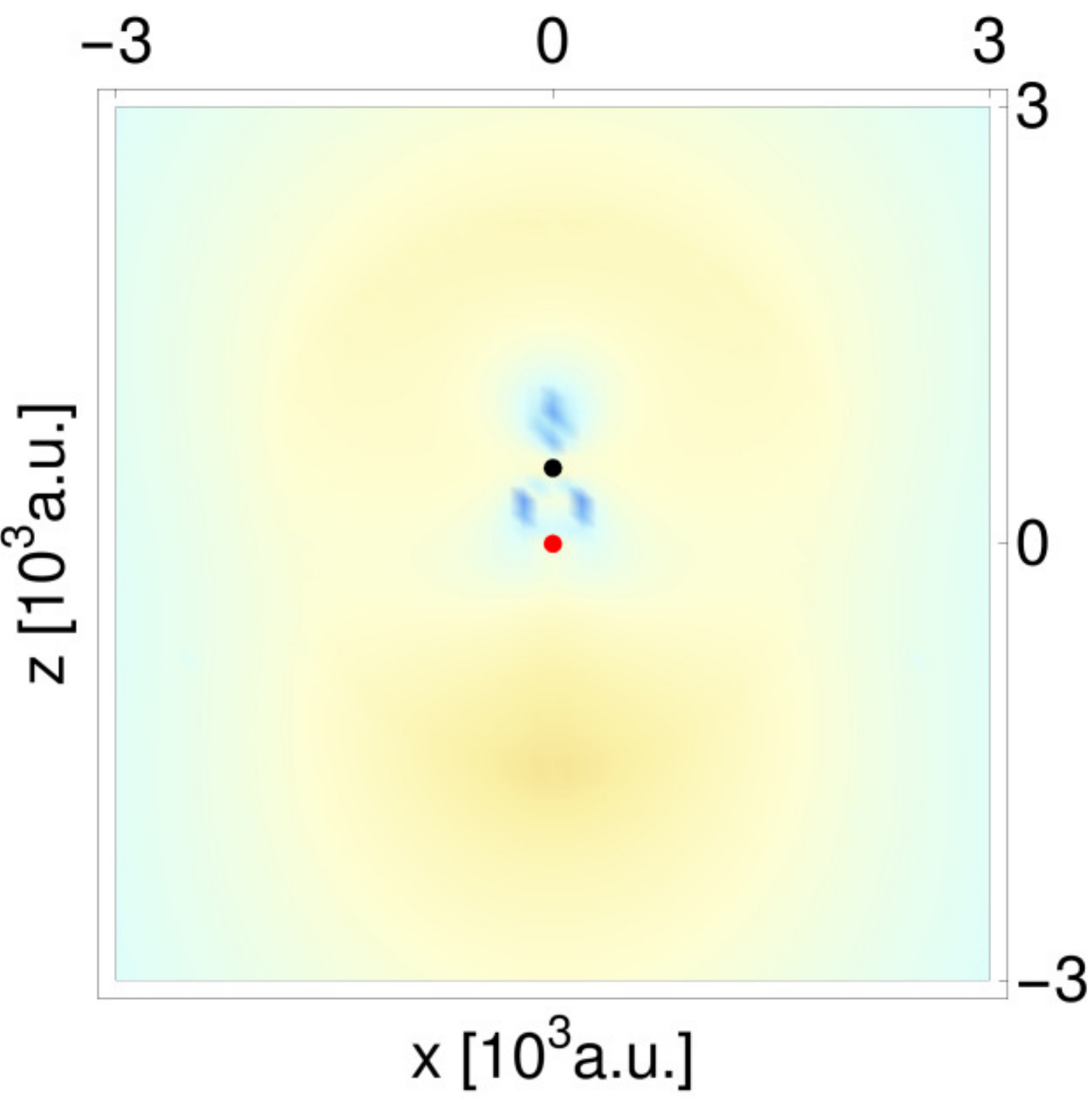}&
\includegraphics[width=.21\textwidth]{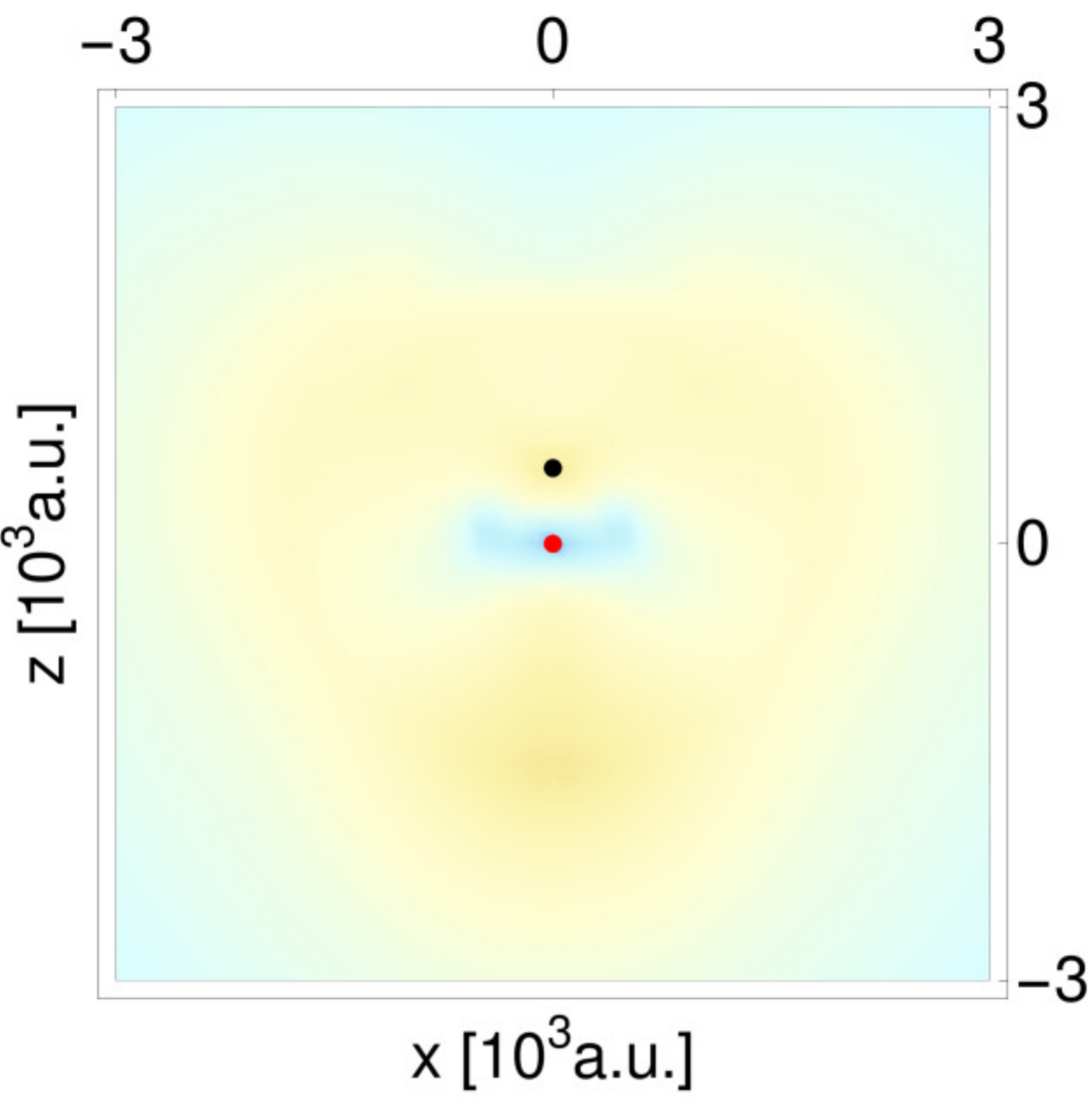}&
\includegraphics[width=.03\textwidth]{Figures/b2_rb.pdf}\\
&(i) & (j) &(k) &(l) & \\
\end{tabular}

\caption{Electrostatic potential $\phi_{LRRM}$ in a. u. generated by a $n=33$ trilobite molecule (first row), a radial butterfly molecule (second row) and a cosine butterfly molecule (third row)in the near region. In (a), (e) and (i)  the exact expression, Eq.(\ref{eq:exa}), is used; in (b), (f) and (j) the monopole term in prolate spheroidal coordinates is used. The relative error, Eq.~(\ref{eq:rel}), in logarithmic scale is illustrated in (c), (g)  and (k) taking up to the spheroidal dipole term in the expansion, and in (d), (h) and (l) taking up to the spherical dipole term. The Rydberg core (red dot) and neutral atom ( black dot) separations are  $\mathsf{r}_\mathrm{t}= 1500$ a$_0$ for the trilobite,  $\mathsf{r}_{\mathrm{rb}}=540$ a$_0$ for the radial butterfly and $\mathsf{r}_{\mathrm{ab}}=520$ a$_0$ for the  cosine butterfly molecule. }\label{fig:near}
\end{figure}

\begin{figure}
\centering
\begin{tabular}{@{}c @{}c @{}c  @{}c  @{}c}
\includegraphics[width=.20\textwidth]{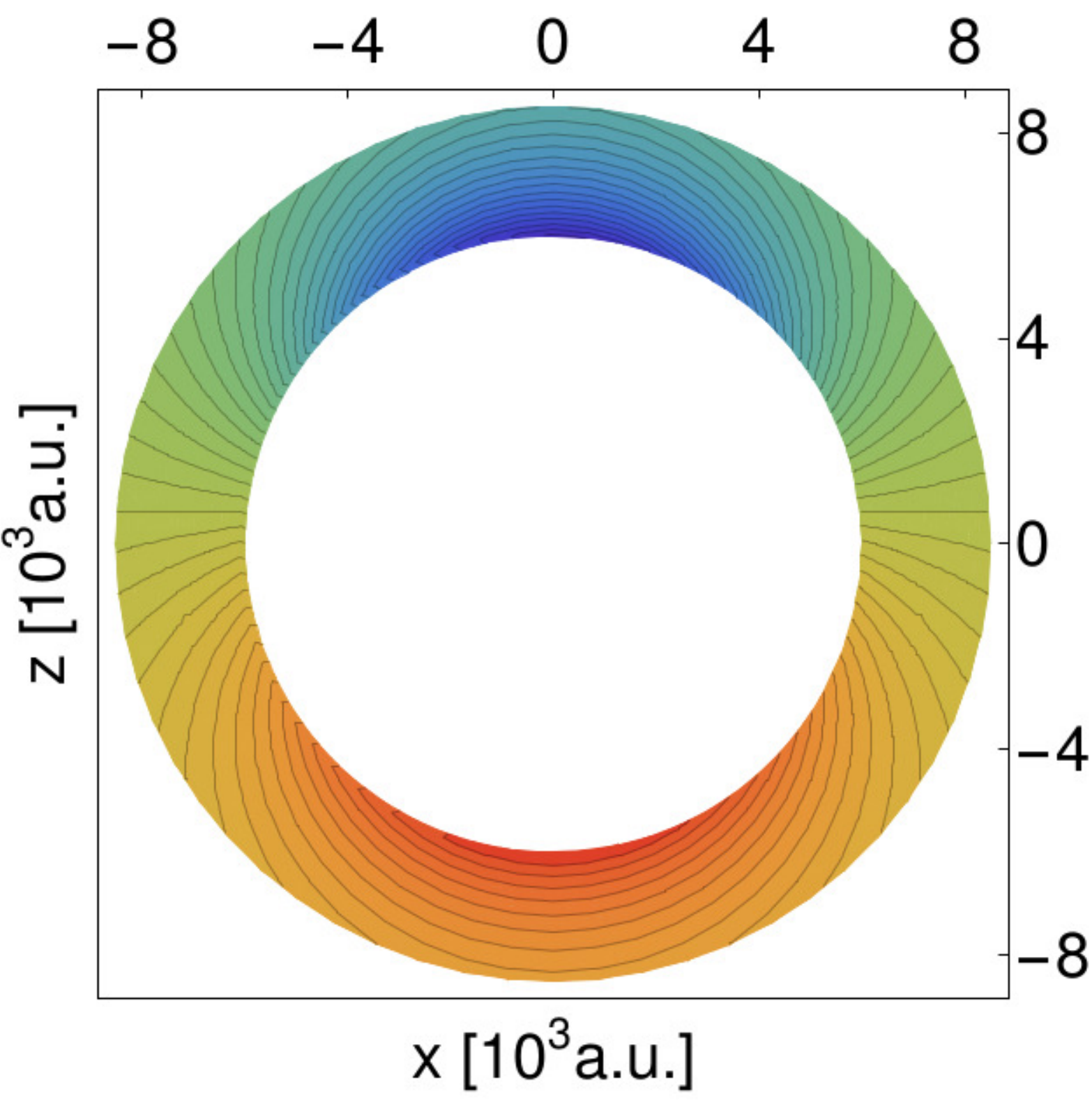}&
\includegraphics[width=.057\textwidth]{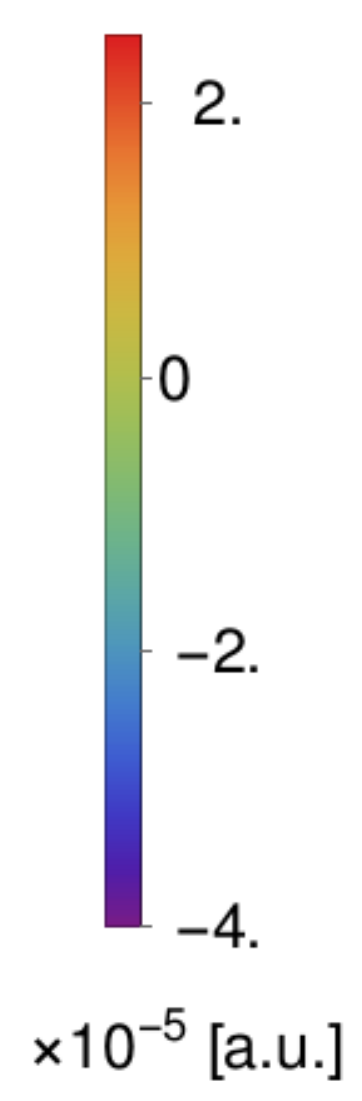}&
\includegraphics[width=.20\textwidth]{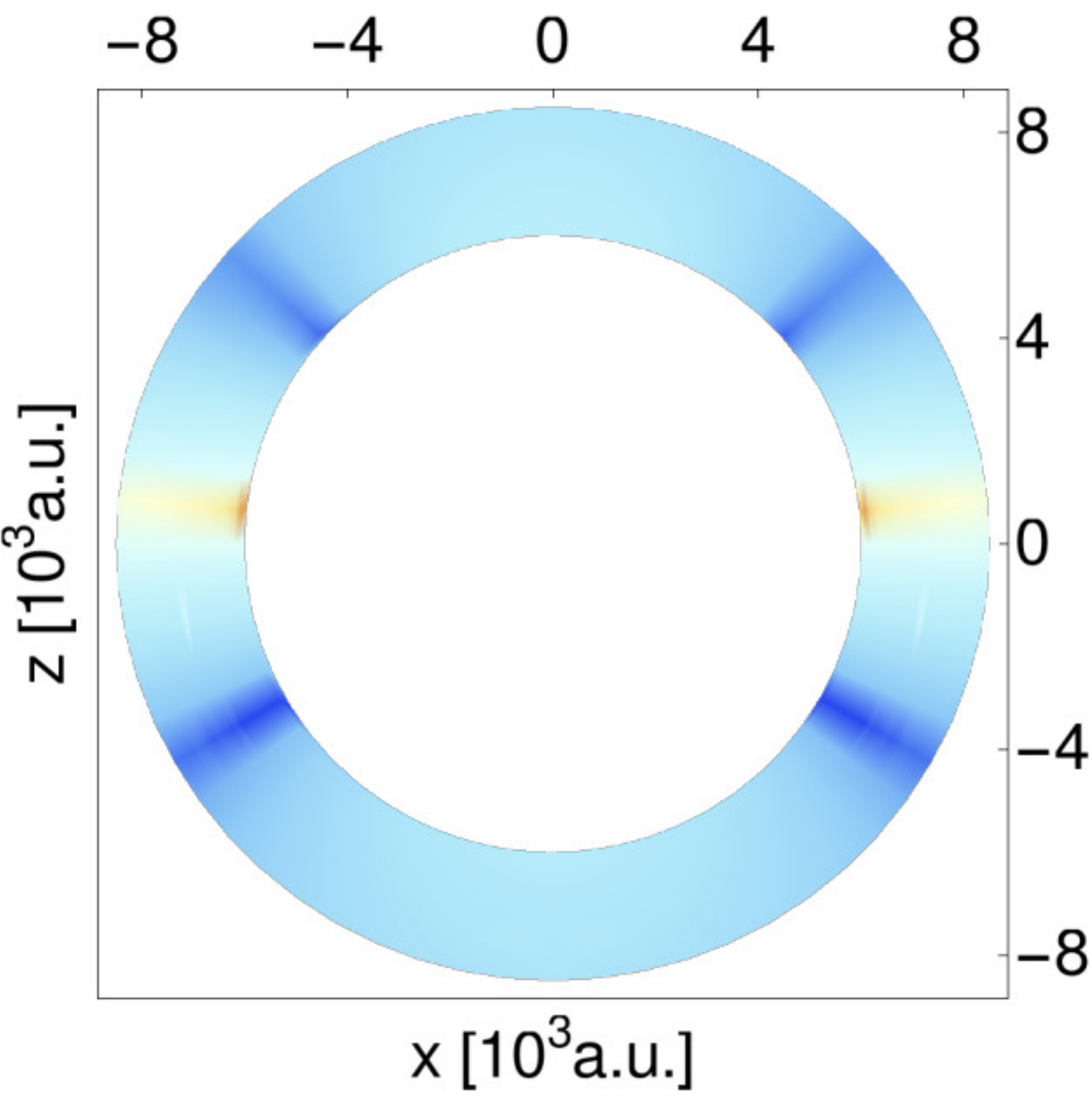}&
\includegraphics[width=.20\textwidth]{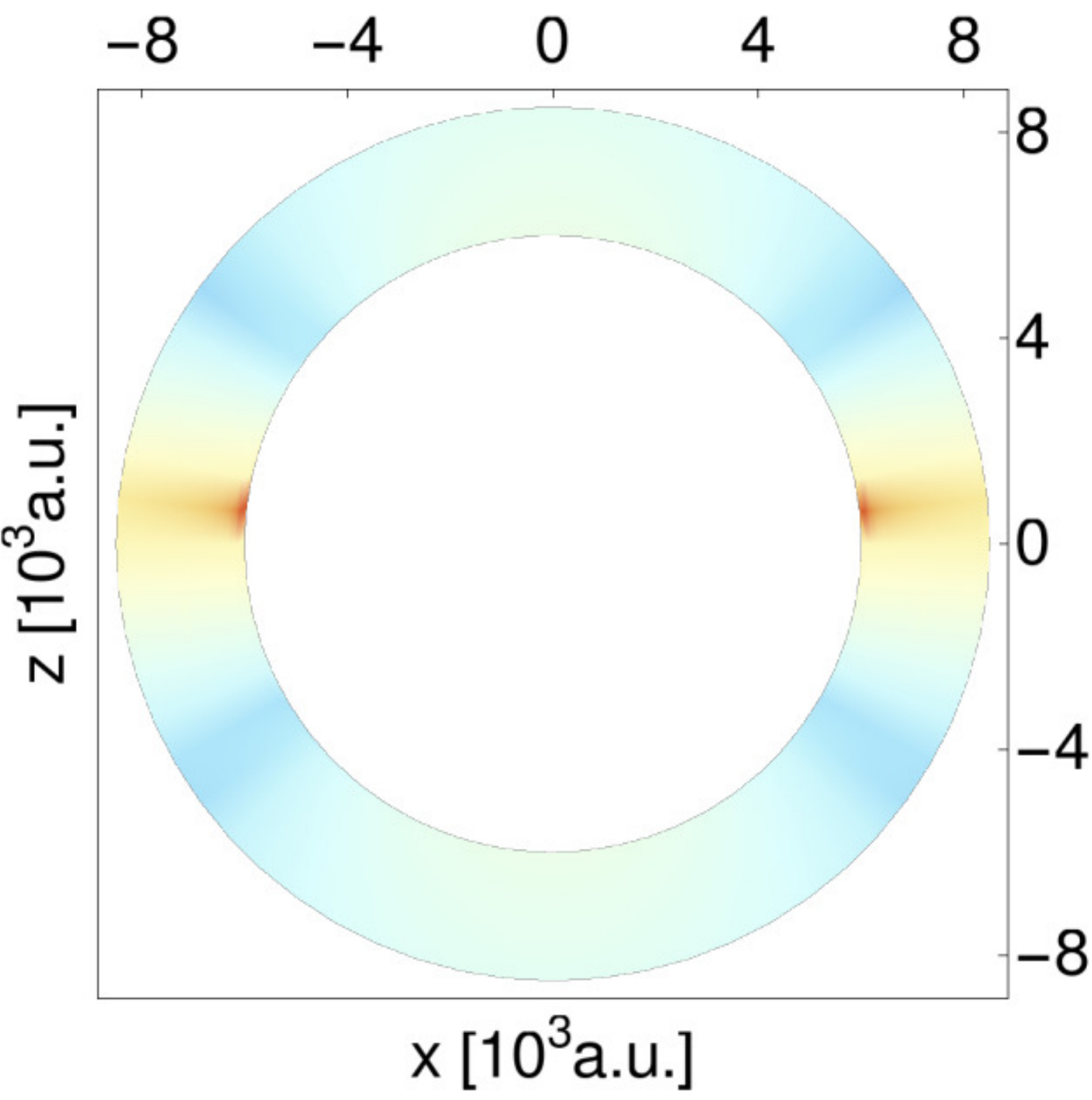}&
\includegraphics[width=.0285\textwidth]{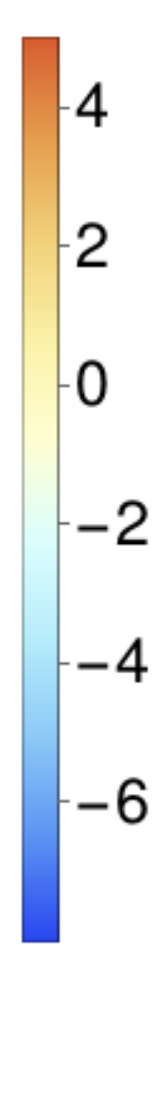}\\
(a) Exact &  & (b) Spheroidal &  (c) Spherical & \\
\includegraphics[width=.20\textwidth]{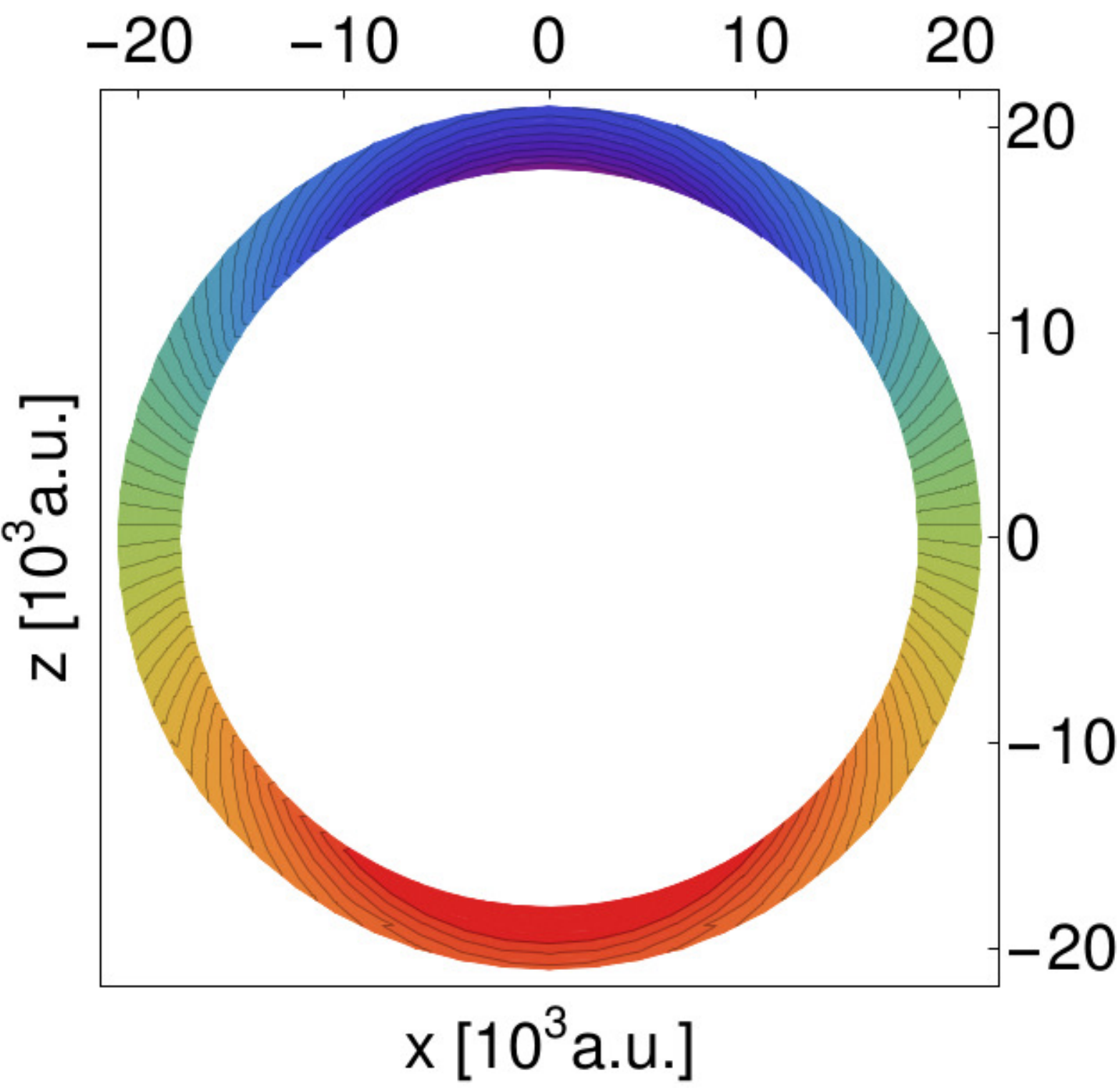}&
\includegraphics[width=.057\textwidth]{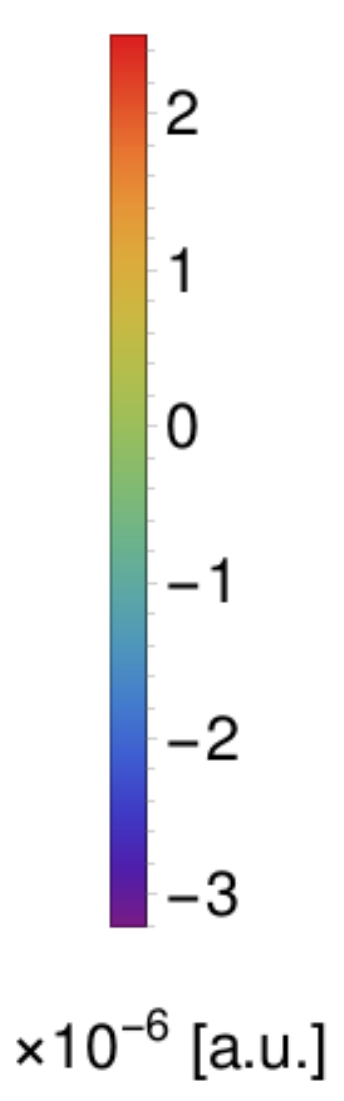}&
\includegraphics[width=.20\textwidth]{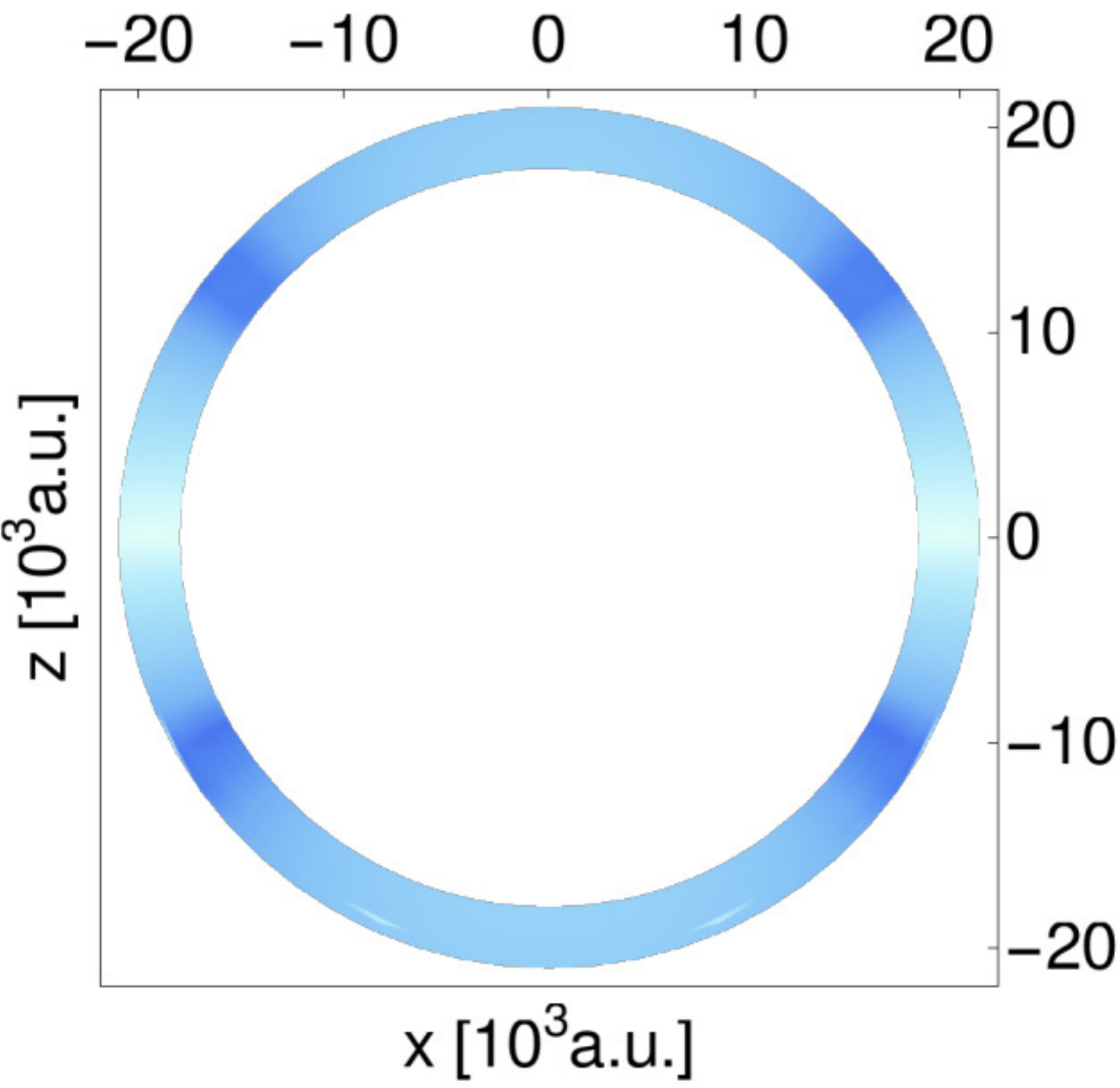}&
\includegraphics[width=.20\textwidth]{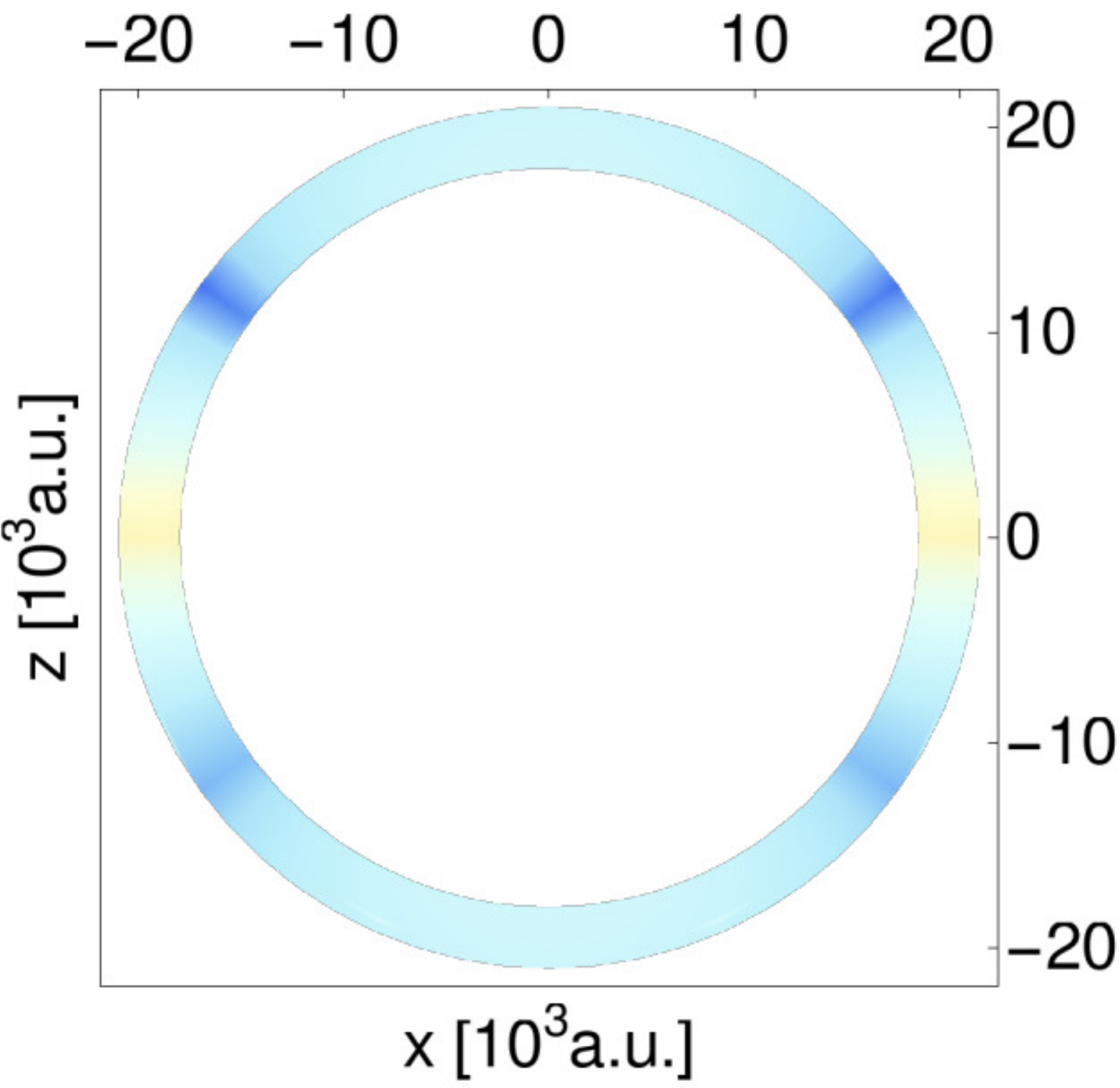}&
\includegraphics[width=.0285\textwidth]{Figures/bd_trilo2.pdf}\\
(d) Exact &  & (e) Spheroidal &  (f) Spherical &  
\end{tabular}
\caption{(a) Electrostatic potential $\phi_{LRRM}$ in a. u. generated by a $n=33$ trilobite molecule in the intermediate region using the exact expression. 
Relative error in logarithmic scale using up to (b) a spheroidal dipole expansion, (c) and a spherical dipole moment expansion. (d) Electrostatic potential for the same molecule in the far region using the exact expression. 
Relative error in logarithmic scale using up to (e) a spheroidal dipole expansion, (f) a spherical dipole moment expansion.
}\label{fig:trilo2b}
\end{figure}

 In Figs.~\ref{fig:near}-\ref{fig:trilo2b}, the electrostatic potentials of longe range Rydberg Rubidium molecules are illustrated at the $xz$ plane. The principal quantum number is fixed, $n=33$, as well as the  nuclear separations $\mathsf{r}$. As in previous illustrative examples, the nuclear separations were determined by the position of a minimum in the  Born-Oppenheimer energy curve of each molecule. In the near region, Fig.~\ref{fig:near}, it can be observed that the monopole description using prolate spheroidal coordinates already reproduces general features of the exact potential. A better visualization of the behavior of the relative error is obtained using a logarithmic scale; for it a smaller relative error corresponds to a more negative value, i. e., the more strong blue is observed in the figure, the better the approximation is. For the trilobite, an expansion that includes up to dipole terms gives a maximum error of  approximately 25 \% in a small region inside the spheroid $\xi\sim 2.2$. On the other hand, for the  trilobite the spherical representation yields a  relative error  of up to 60 \% close to the neutral atom and the region in which the error is $\sim$ 25\% involves
extended zones in the square defined by $-\mathsf{r}_\mathrm{t}<x<\mathsf{r}_\mathrm{t}$ and $-\mathsf{r}_\mathrm{t}<z<\mathsf{r}_\mathrm{t}$.
In the second row of Fig.~\ref{fig:near} the 
electrostatic potential of the radial butterfly molecule in the near region is shown; there is a closer resemblance to the exact potential of both the spheroidal and spherical dipole approximations. This could be expected from the density distribution of the electronic orbital, Fig.~\ref{fig:regiones}. Nevertheless the spheroidal description is more accurate than the spherical one. Finally, in the third row 
the cosine butterfly potential representation  up to dipole terms is shown. It is
observed that this representation does not reproduce accurately the exact potential at the region $-2 \mathsf{r}_{\mathrm{ab}}<z< -5\mathsf{r}_{\mathrm{ab}}$ both in the spherical and the spheroidal multipole expansions. This indicates an increased relevance of higher multipole terms in the description of cosine butterfly molecules. As expected from the evaluation of the momenta of the cosine butterfly LRRM, quadrupole moments play a more important  role in this case than in the trilobite or radial butterfly configuration.

In Fig.~\ref{fig:trilo2b},  the exact potential of the same $n =33$ trilobite molecule is illustrated for the intermediate region (ring defined by $4\mathsf{r}_\mathrm{t} =6000\, \mathrm{a}_0<\sqrt{x^2 + z^2}<8500\,\mathrm{a}_0 =5.7\mathsf{r}_\mathrm{t}$), and for the far region (ring defined by $10.66\mathsf{r}_\mathrm{t}= 16000\, \mathrm{a}_0<\sqrt{x^2 + z^2}<20000\, \mathrm{a}_0= 13.33\mathsf{r}_\mathrm{t}$).  Observe that in the far region the spherical dipole approximation exhibits a less deficient representation of the exact potential; nevertheless,  in the three regions, the spheroidal dipole approximation gives a better representation of the electrostatic potential than the spherical one.
Similar  results are obtained from an analysis of the butterfly LRRM molecules.

It is remarkable that for the three kind of molecules the monopole term already provides significant information about the electrostatic field which is encoded in the appropriate selection of the vector $\vec{\mathsf{r}}$ that defines the coordinate system. In fact, in spite of the similarity between the dipole and quadrupole dimensionless multipole moments, for the trilobite and radial butterfly molecule, the general behavior of the potential even at short distances is well reproduced by the spheroidal multipole expansion up to the dipolar term,
\begin{equation}
\phi^{(t,rb)}_{LRRM}(\vec{r}) \approx\frac{1}{r}+ \frac{2}{\mathsf{r}}\Big[\widetilde{\mathcal{T}}_{00} \, V_{00}^e(\vec r)+ 3 \Big[\widetilde{\mathcal{T}}_{10} \,  V_{10}^e(\vec r)-\frac{1}{2}\Big(\frac{1}{\sqrt{2}} \widetilde{\mathcal{T}}_{1-1} \,  V_{11}^e(\vec r) 
+2\sqrt{2} \widetilde{\mathcal{T}}_{11} \,  V_{1-1}^e(\vec r)\Big)\Big]\Big].\label{eq:far}
\end{equation}
Meanwhile, for the cosine butterfly molecule, quadrupole contributions must be included
\begin{equation}
\phi^{(cb)}_{LRRM}(\vec{r}) \approx\frac{1}{r}+ \frac{2}{\mathsf{r}}\Big[\widetilde{\mathcal{T}}_{00} \, V_{00}^e(\vec r)+ 3 \Big[\widetilde{\mathcal{T}}_{10} \,  V_{10}^e(\vec r)-\frac{5}{2}\Big( \widetilde{\mathcal{T}}_{20} \,  V_{20}^e(\vec r) 
+\frac{1}{24\sqrt{6}} \widetilde{\mathcal{T}}_{22} \,  V_{22}^e(\vec r)
+ 4\sqrt{6} \widetilde{\mathcal{T}}_{2-2} \,  V_{2-2}^e(\vec r)
\Big)\Big]\Big].\label{eq:far:ab}
\end{equation}

\section{LRRM intermolecular interactions.}
Intermolecular interactions in quantum chemistry are usually studied considering each molecule as an approximate localized charge distribution to which an internal multipole characterization is performed. The permanent or induced multipoles are assumed to interact according to the spherical multipole expansion of the Coulomb potential.
 The electrostatic potential between a pair of molecules with charge densities $\rho_A(\vec r)$ and $\rho_B(\vec r)$ measured with respect to
selected points within each molecule which are separated by a vector $\vec R$ is given by
\begin{equation}
\mathcal{V} (\vec R; A; B) =  \int d^3r \int d^3 r^\prime\frac{\rho_A(\vec r)\rho_B(\vec r^\prime)}{\vert \vec r - \vec r^\prime - \vec R\vert}.
\end{equation}
If $\vert \vec R\vert >>\vert \vec r - \vec r^\prime \vert$ for all the vectors $\vec r$ and $\vec r^\prime$ at
which $\rho_A(\vec r)$ and $\rho_B(\vec r^\prime)$ are not negligible, that is, if the constituient molecules A and B are far enough, an {\it external} multipole expansion gives reliable results. Then, it can be shown \cite{stone} that in the spherical multipole approach 
\begin{align}
\nonumber \mathcal{V}(\vec R; A; B) \approx&\sum_{\ell_1, \ell_2} \sum_{k_1,k_2} (-1)^{\ell_1} \sqrt{\frac{(2 \ell_1+2\ell_2+1)!}{(2\ell_1)!(2 \ell_2)!}} {T}^{s*}_{\ell_1 k_1}(A) {T}^{s*}_{\ell_2 k_2}(B) \\
&\times \sum_{m_1, m_2, m}  \left[ D_{m_1 k_1}^{\ell_1}(\Omega_A) \right]^*   \left[ D_{m_2 k_2}^{\ell_2}(\Omega_B) \right]^*  
\begin{pmatrix}
\ell_1 & \ell_2 & \ell_1+\ell_2 \\
m_1 & m_2 & m
\end{pmatrix}
\mathcal{J}_{\ell_1+\ell_2,m}(\vec{R}). \label{eq:ms}
\end{align}
This expression involves the spherical multipole-multipole interaction term $\mathcal{J}_{\ell,m}(\vec{R})$,
\begin{equation}
\mathcal{J}_{\ell,m}(\vec{R}) =\sqrt{\frac{(\ell-m)!}{(\ell +m)!}} U^e_{\ell m}(\vec R),
\end{equation}
the spherical multipole moments
\begin{equation}
{T}^s_{\ell m}=\sqrt{\frac{(\ell-m)!}{(\ell +m)!}}\int d^3r\rho(\vec r) U^i_{\ell m}(\vec r),
\end{equation}
and the Wigner matrices $D_{m k}^{\ell}(\Omega)$; the presence of the latter is due to the reorientation, via a rotation with Euler angles $\Omega=(\alpha,\beta,\gamma)$, of the reference frame used in the evaluation of the spherical multipole moments of a given molecule to that used to describe the binary molecule system as a whole. 

The external spheroidal multipole expression of $\mathcal{V} (\vec R; A; B)$ has a similar structure to Eq.~(\ref{eq:ms}), though it  incorporates the existence of a natural scale factor $\mathsf{r}$ for each molecule. To obtain it, Eq.~(\ref{eq:multipole}) is applied, and then usage is made of the behavior of the internal spheroidal functions $V_{\ell m}^i(\vec r)$ under translations, rotations and scaling transformations~\cite{jansen2000}; the latter are summarized in the Appendix.
The result is
\begin{eqnarray}
\mathcal{V}(\vec R; A; B) &\approx&  \sum_{\ell} \sum_{m} \tilde{\mathcal{V}}_{\ell m}(\vec R; A; B)\nonumber\\ 
&=& \sum_{\ell} \sum_{m} \mathcal{I}^*_{\ell,m}(\vec{R};\vec{\mathsf{r}}_A)\Big[\sum_{j_1,j_2} \sum_{m_1,m_2} \left(\frac{\mathsf{r}_A}{2}\right)^{\ell-j_1-j_2}(-1)^{j_2} t_{m_1 m_2 m}^{j_1j_2 \ell} \mathcal{T}_{j_1 m_1}(A; \vec{\mathsf{r}}_A)\nonumber\\
&\cdot&
 \sum_{k}\sum_{m^\prime}\Big(\frac{\mathsf{r}_A}{2}\Big)^{j_2 -k}\mathcal{D}_{m^\prime m_2}^{k j_2}(\Omega_{AB})
\sum_{k^\prime}\Big(\frac{\mathsf{r}_A}{2}\Big)^{k -k^\prime}{\mathcal{S}}^{k k^\prime}_{m^\prime}(\mathsf{r}_B/\mathsf{r}_A)\mathcal{T}_{k^\prime m^\prime}(B; \vec{\mathsf{r}}_B)\Big]
	\label{eq:transfo1}
	\end{eqnarray} 
In this equation, we have chosen to describe the dimolecular system using the  reference frame that corresponds to the isolated A molecule: the multipole moment of the molecule $B$, $\mathcal {T}^\prime _ {j m} (B; \vec{\mathsf{r}} _A)$, evaluated in the spheroidal coordinate system defined by $\vec{\mathsf{r}} _A$, must be expressed in terms of the multipoles $\mathcal{T} _ {j_1 m_1} (B; \vec{\mathsf{r}} _B)$ in the spheroidal coordinate system used initially to describe the $B$ molecule. This required a rotation by Euler angles $\Omega_{AB}$ incorporated via the generalized Wigner matrices $\mathcal{D}_{m^\prime m_2}^{k j_2}(\Omega)$.
Notice that this expression assumes that the internal multipole moments $\mathcal {T}_ {\ell m} (C; \vec{\mathsf{r}} _C)$ were evaluated with respect to the Rydberg core of molecule $C$ with $C=A,B$. If a different origin of coordinates were chosen, the appropriate translation should be performed according to Eq.~(\ref{eq:translation}) that is given in the Appendix.

In Eq.~(\ref{eq:transfo1}), the indices $\ell$ and $m$ define the multipole effective potential
$\mathcal{I}^*_{\ell,m}(\vec{R};\vec{\mathsf{r}}_A)$. In the same equation, the coefficients
$t_{m_1 m_2 m}^{j_1j_2 \ell}$ may be different from zero only if
$j_1+j_2+\ell$ is even, $m_1+m_2 = m$ and $\ell\ge j_1+j_2\ge\vert m\vert$. So that the internal multipole terms $j_1$ and $j_2$, that go along the $\ell$ external multipole potential $\mathcal{I}^*_{\ell,m}$ satisfy also such conditions.

The simplest scenario for a multipole description considers the atomic systems far apart so that, besides the external
multipole expansions can be applied, the electronic orbitals in a given molecule are not modified within  a first order approximation by the presence of the other molecules. Corrections to this perturbative scheme are applied {\it a posteriori}. Consider   a pair of LRRM molecules $A$ and $B$  with a vector $\vec R$  joining the two Rydberg nuclei, and  $\vec{\mathsf{r}}_A$ and $\vec{\mathsf{r}}_{B}$ the vectors that join the Rydberg and neutral atom in each molecule. If $ R \gg \mathsf{r}_A, \mathsf{r}_B$, the interaction energy derived by the effective scattering of electron in A(B) by the neutral atom in B(A) is negligible, as well as the contribution to the intermolecular interaction due to electronic exchange.  Within these approximations the electrostatic energy between a pair of LRRM molecules $A$ and $B$  is
\begin{equation}
\mathcal{V} (\vec R; A; B) =\frac{1}{R} -\int d^3r \frac{\left| \Upsilon_A(\vec{r}; \vec{\mathsf{r}}_A) \right|^2}{\vert \vec r - \vec R\vert} -\int d^3r \frac{\left| \Upsilon_B(\vec{r}; \vec{\mathsf{r}}_B) \right|^2}{\vert \vec r + \vec R\vert}
+ \int d^3r \int d^3 r^\prime\frac{\left| \Upsilon_A(\vec{r}; \vec{\mathsf{r}}_A) \right|^2\left| \Upsilon_B(\vec{r}^\prime; \vec{\mathsf{r}}_B) \right|^2}{\vert \vec r^ - \vec r^\prime - \vec R\vert}.
\end{equation}

\subsection{ LRRM with parallel orientation of their natural axes.}

\begin{figure}[t]
	\begin{center}
		\includegraphics[width=.35\textwidth]{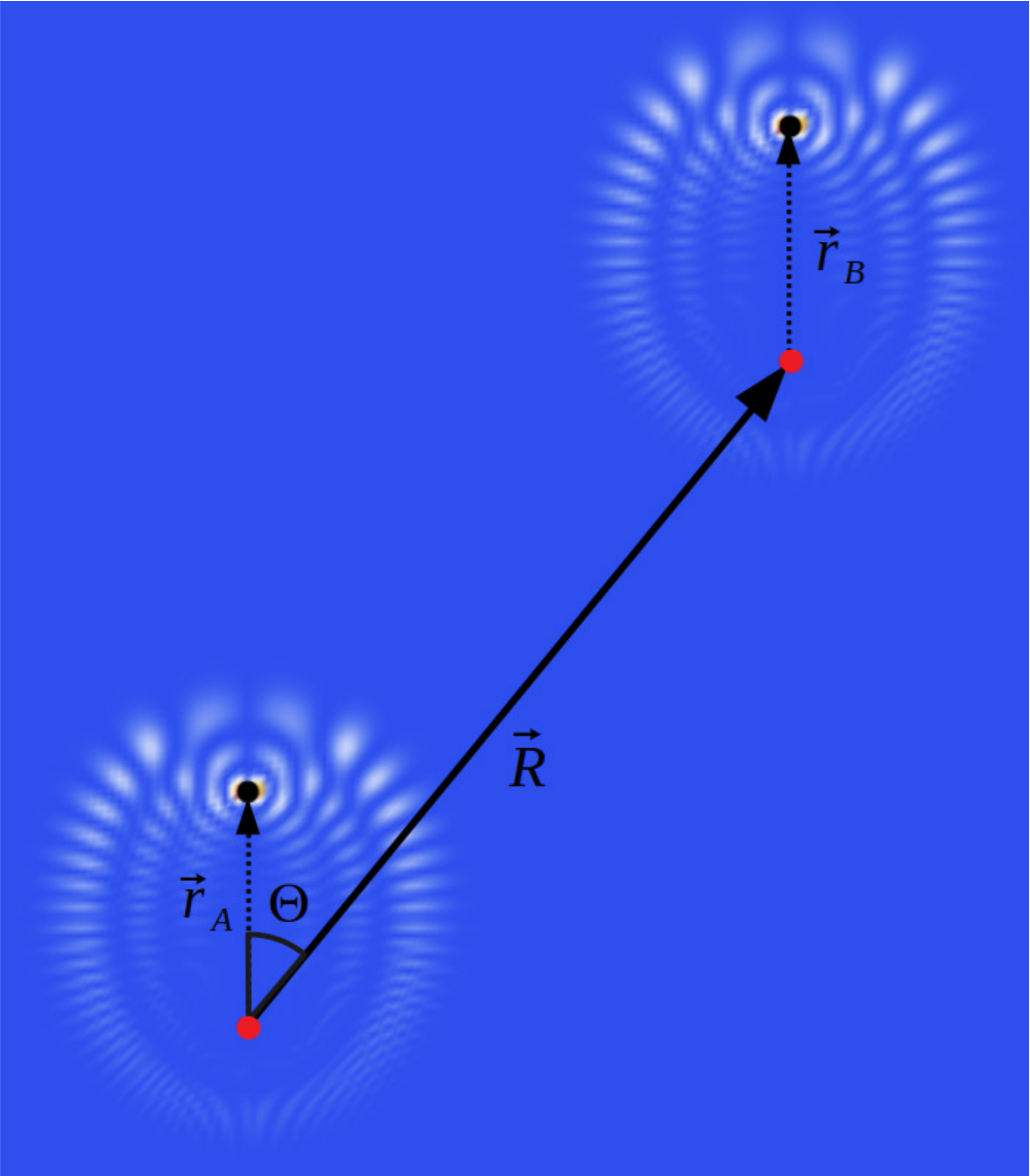}
		\caption{Coordinates  used to describe the interaction energy for two LRRM with parallel orientation of their natural axes. For $ \vec {\mathsf{r}}_A\times \vec {\mathsf{r}}_B=\vec 0$, it is possible to choose a plane where $\vec{\mathsf{r}}_A$ and $\vec {\mathsf{r}}_B$ are contained. The intermolecular vector $\vec R$ that joins the two Rydberg cores is also contained in the same plane and makes an angle $\Theta$ with $\vec{\mathsf{r}}_A$.  }
		\label{fig:angulo}
	\end{center}
\end{figure}

The  spherical and spheroidal dipole internal moments of a given molecule coincide. For trilobite and butterfly molecules these moments are parallel to the internuclear axis $\vec{\mathsf{r}}$. As a consequence, long range diatomic molecules in the presence of a  homogeneous external electric field will orient their internal vector  $\vec{\mathsf{r}}$  along that field. If the field is weak enough, the
electronic orbitals can be approximated by their expressions in the absence of that external field.
Experiments along these lines have already been performed \cite{niederprum2016}. In this subsection,
we obtain the expectations from calculations using spheroidal multipole expressions. Emphasis will be given in the similarities and differences between standard (spherical) dipolar studies and spheroidal ones.

We have numerically studied the convergence of the spherical and spheroidal multipole expansions. The illustrative example reported in this Section corresponds to  the trilobite and cosine  butterfly LRRM  molecules with $n=33$ whose 
electronic densities are shown in Fig.~\ref{fig:regiones}. The general expectations for radial butterfly molecules have already been reported in details at Ref.~\cite{eiles2017b}

\subsubsection{Trilobite molecules.}
For trilobite and radial butterfly LRRM with parallel  internal axis $\vec{\mathsf{r}}_A$ and $\vec{\mathsf{r}}_B$, the interaction energy depends only on the separation distance $R$ and  the angle $\Theta$ between $\vec{R}$ and $\vec{\mathsf{r}}_A$ as illustrated in Fig.~\ref{fig:angulo}; it is  independent on the azimuthal angle $\Phi$.
The origin of the spheroidal coordinates is located at the center of the line that joins the Rydberg core and the neutral atom. The vector $\vec R$ joins the pair of Rydberg cores located at each LRRM. This must be taken into account in the evaluation of the multipole expansion.
It results that
\begin{equation}
\int \frac{| \Upsilon_A( \vec{r};\vec{\mathsf{r}}_A)|^2}{|\vec{r}+\frac{\vec{\mathsf{r}}_A}{2}-\vec{R}|} d^3r = \frac{2}{\mathsf{r}_A} \sum_{\ell=0}^{\infty} G_{\ell}(A;\vec{\mathsf{r}}_A) \tilde{\mathcal{I}}_{\ell 0}(\vec{R};\vec{\mathsf{r}}_A),
\label{eq:parA}
\end{equation}
and
$$\iint \frac{| \Upsilon_A( \vec{r};\vec{\mathsf{r}}_A)|^2 | \Upsilon_B( \vec{r}';\vec{\mathsf{r}}_B)|^2}{|\vec{r}+\frac{\vec{\mathsf{r}}_A}{2}-\frac{\vec{\mathsf{r}}_B}{2}-\vec{r'}-\vec{R}|} d^3r d^3r^\prime=$$
\begin{equation}
 \frac{2}{{\mathsf{r}}_A} \sum_{\ell=0}^{\infty} \left[ \sum_{j_1,j_2} (-1)^{j_2}  t_{0 0 0}^{j_1 j_2 \ell} G_{j_1}(A;\vec{\mathsf{r}}_A) \left( \sum_{k=0}^{j_2} \left( \frac{{\mathsf{r}}_B}{{\mathsf{r}}_A} \right)^{k} S_0^{j_2 k}\left( \frac{{\mathsf{r}}_B}{{\mathsf{r}}_A} \right) G_k(B;\vec{\mathsf{r}}_B) \right) \right] \tilde{\mathcal{I}}_{\ell 0}(\vec{R};\vec{\mathsf{r}}_A)
\label{eq:parAB}
\end{equation}
with
\begin{equation}
G_k(A;\vec{\mathsf{r}}_A)= \sum_{j_1,j_2}   t_{0 0 0}^{j_1 j_2 k} \frac{j_2!}{(2 j_2-1)!!} \tilde{\mathcal{T}}_{j_1 0}(A;\vec{\mathsf{r}}_A). 
\end{equation}

The multipole interaction potentials $\tilde{\mathcal{I}}_{\ell,m}(\vec{R};\vec{\mathsf{r}}_A)$ depend on the
spheroidal coordinates which also depend on  $\vec{R}$ through its modulus $R=\left|\vec{R}\right|$ and the angle $\Theta$ between $\vec{R}$ and $\vec{\mathsf{r}}_A$.
For identical molecules with parallel symmetry axes, the coefficient of $\tilde{\mathcal{I}}_{\ell,m}(\vec{R};\vec{\mathsf{r}}_A)$ for $\ell$ odd in Eq.~(\ref{eq:parAB}) is null; and a similar result applies to the spherical expansion. 

\begin{figure}
	\begin{center}
		\includegraphics[width=.5\textwidth]{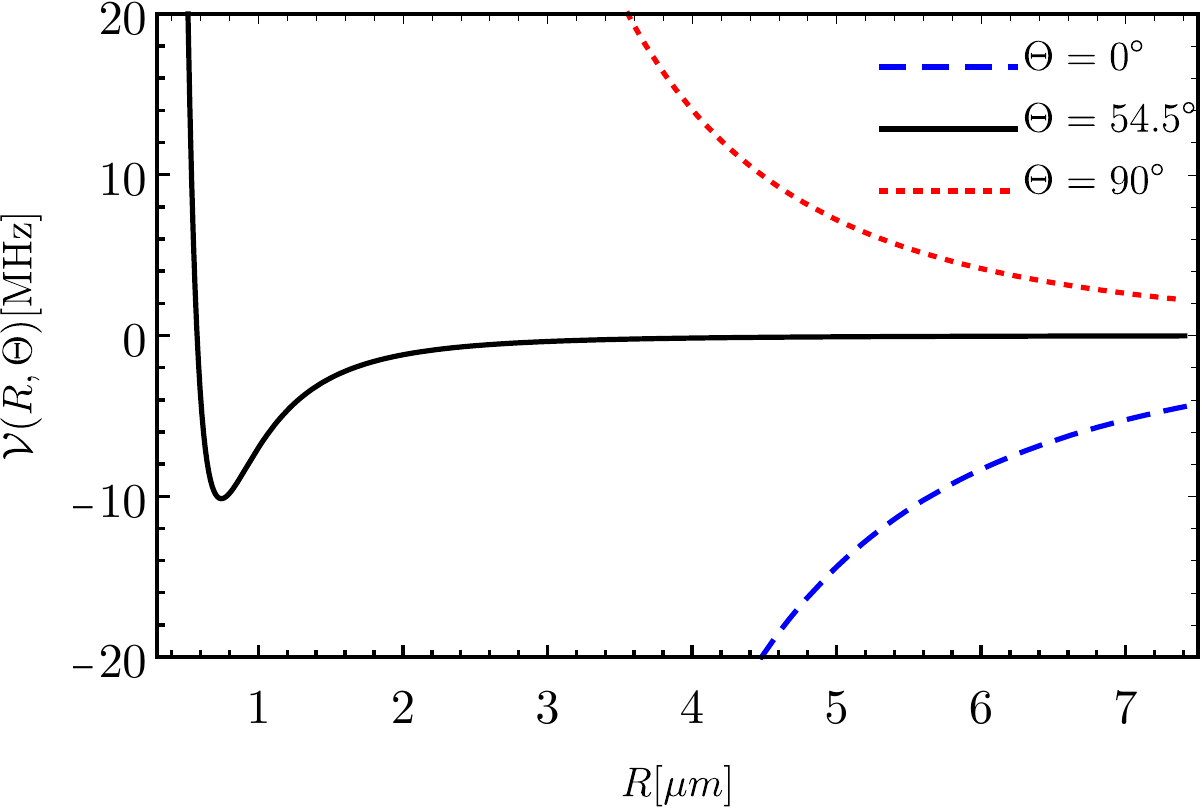}
		\caption{Converged intermolecular potential $\sum_{\ell\le 10} \mathcal{V}_{\ell 0}(R,\Theta)$ for a pair of trilobite $n=33$ molecules for three illustrative angles. \label{par:3angles}}
	\end{center}
\end{figure}

For a pair of $n=33$ Rb trilobite molecules with $\mathsf{r}_A=\mathsf{r}_B = 1500 \mathrm{a}_0$, the partial sums for the total interaction energy  up to $\ell=10$ were calculated. It was also checked that, within six significant figures the contribution of terms with  $\ell>10$ was negligible. 
Since the axes of the molecules are parallel, we considered $\Theta\; \in\; [0^\circ , 90^\circ]$, since for $\Theta \;\in\; [90^\circ , 180^\circ]$, the results are the same that with the supplementary angle $180^\circ -\Theta$.
Figure \ref{par:3angles} illustrates the converged intermolecular potential $\mathcal{V}_{\ell=10}$ for three representative angles. It is found that nearby the so called "magic" angle $\Theta_m\approx 54.73^\circ$, defined by $P_{2}^0(\cos \Theta_m)) = 0$, a minimum in the intermolecular curve is located. In the case of a pair of radial butterfly molecules this angle 
exhibits a local maxima that has been interpreted in terms of a blockade phenomenon \cite{eiles2017b}. At this angle in the spherical multipole expansion just $\ell>2$ multipole interaction terms contribute to the intermolecular interaction.

\begin{figure}
\centering
\begin{tabular}{@{}c @{}c}
	\includegraphics[width=0.45\textwidth]{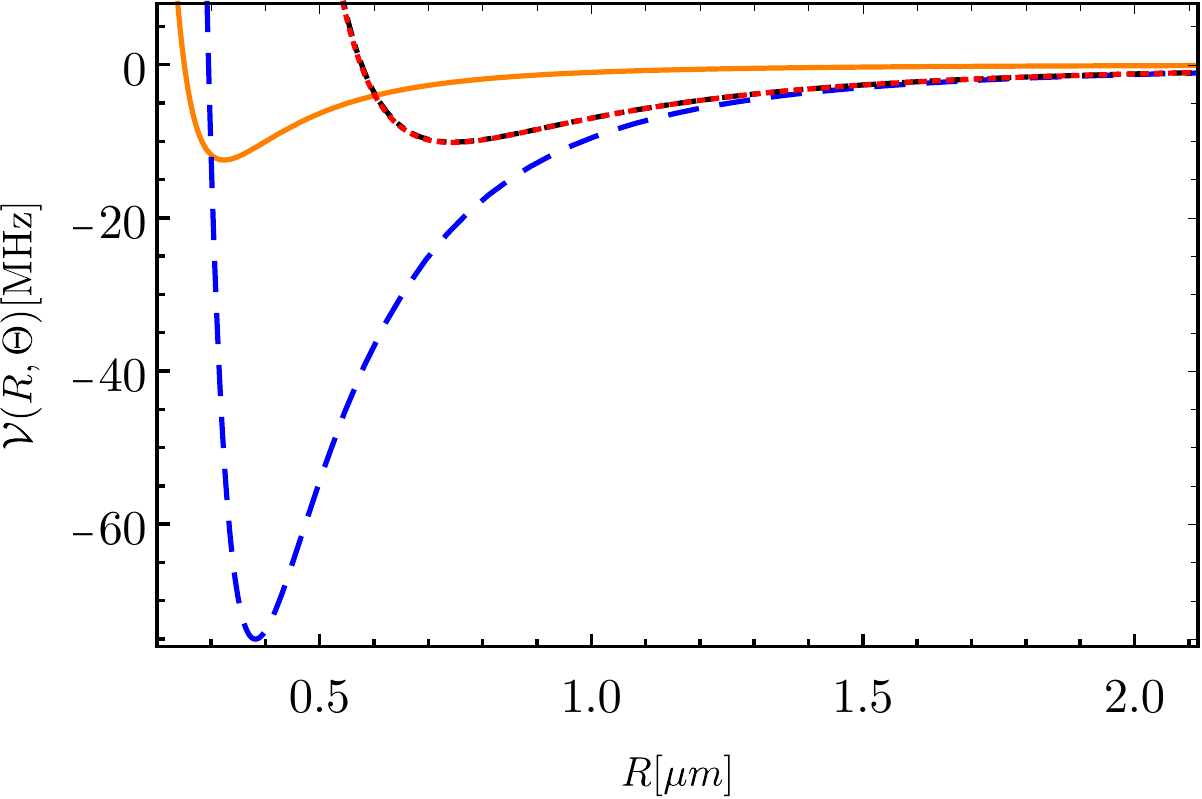} & \includegraphics[width=0.45\textwidth]{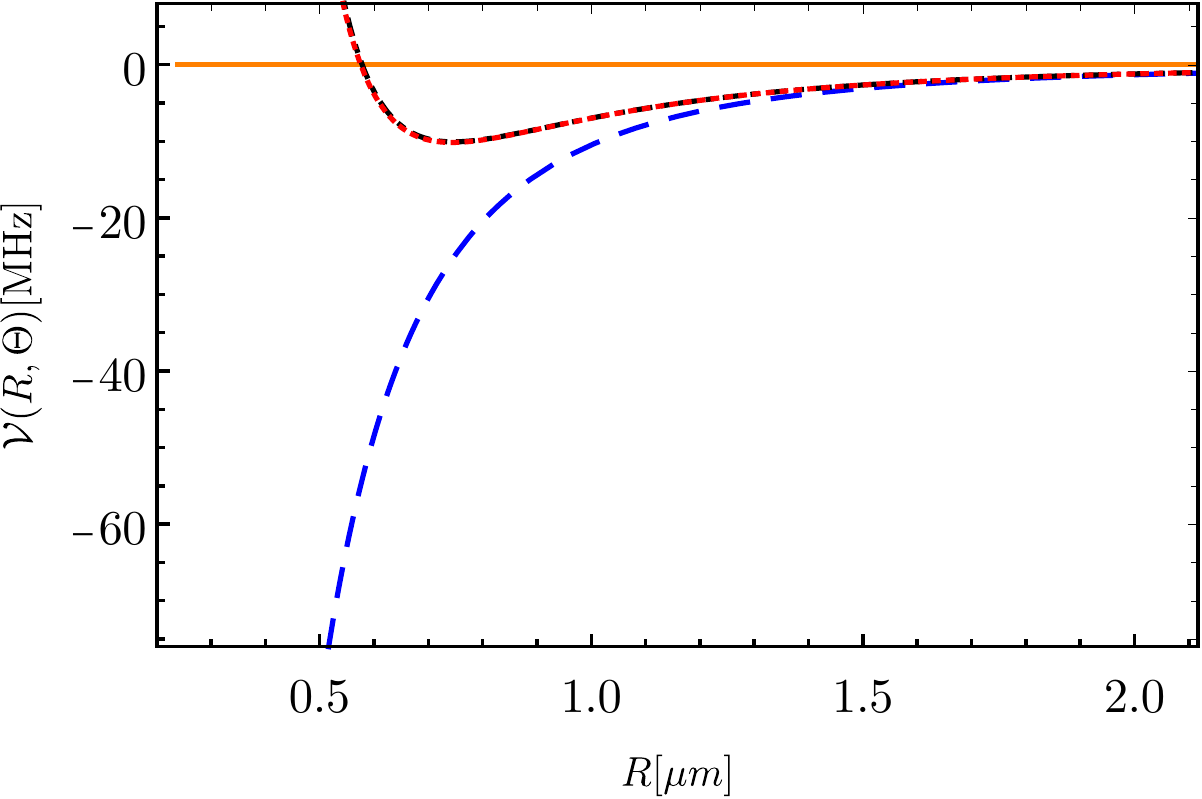}\cr
(a) Spheroidal expansion & (b) Spherical expansion
		
\end{tabular}
		\caption{Convergence of the multipole expansion for a pair of $n=33$ Rb trilobite molecules in the parallel configuration, $\Theta = 54.5^\circ$;
$\ell=0$ corresponds to solid orange line; $\ell=2$  to dashed blue line; $\ell=4$ to red dotted line; $\ell=10$ to black dot dashed line.}
		\label{fig:c1}
\end{figure}

Figure \ref{fig:c1} gives an illustrative comparison between the spheroidal and spherical expansions for $\Theta=54.5^\circ$. It is observed that when multipole terms up to $\ell= 4$ are included both spheroidal and spherical results yield similar results. Notice that for the case of identical LRRM with parallel axes, the multipole interaction term  $\ell$ involves the individual multipole moments $\ell_A$ and $\ell_B$ with
$\ell\ge \ell_A +\ell_B$. For instance, in the $\ell = 2$ term the monopole-monopole, monopole-dipole, dipole-dipole, and  monopole-quadrupole momenta contributions are involved. For the spherical expansion  smaller values of $\ell$ always predict an incorrect qualitative behavior of the intermolecular interaction; meanwhile, the spheroidal expansion even within the monopole interaction ($\ell=0$) predicts a global minimum  at $\Theta = 54.5^\circ$. Nevertheless, the correct localization and depth of the minimum requires including higher multipole contributions. It must be mentioned that for angles that are not in the neighbourhood of the magic angle,  the predicted qualitative behavior of  $\mathcal{V}_{\ell}(\vec{R})$ from the  spheroidal and spherical expansions are the same. However, for all values of $\Theta$, the convergence of the spheroidal multipole expansion is faster than its spherical analogue.

\begin{figure}[H]
	\begin{center}
		\includegraphics[width=.5\textwidth]{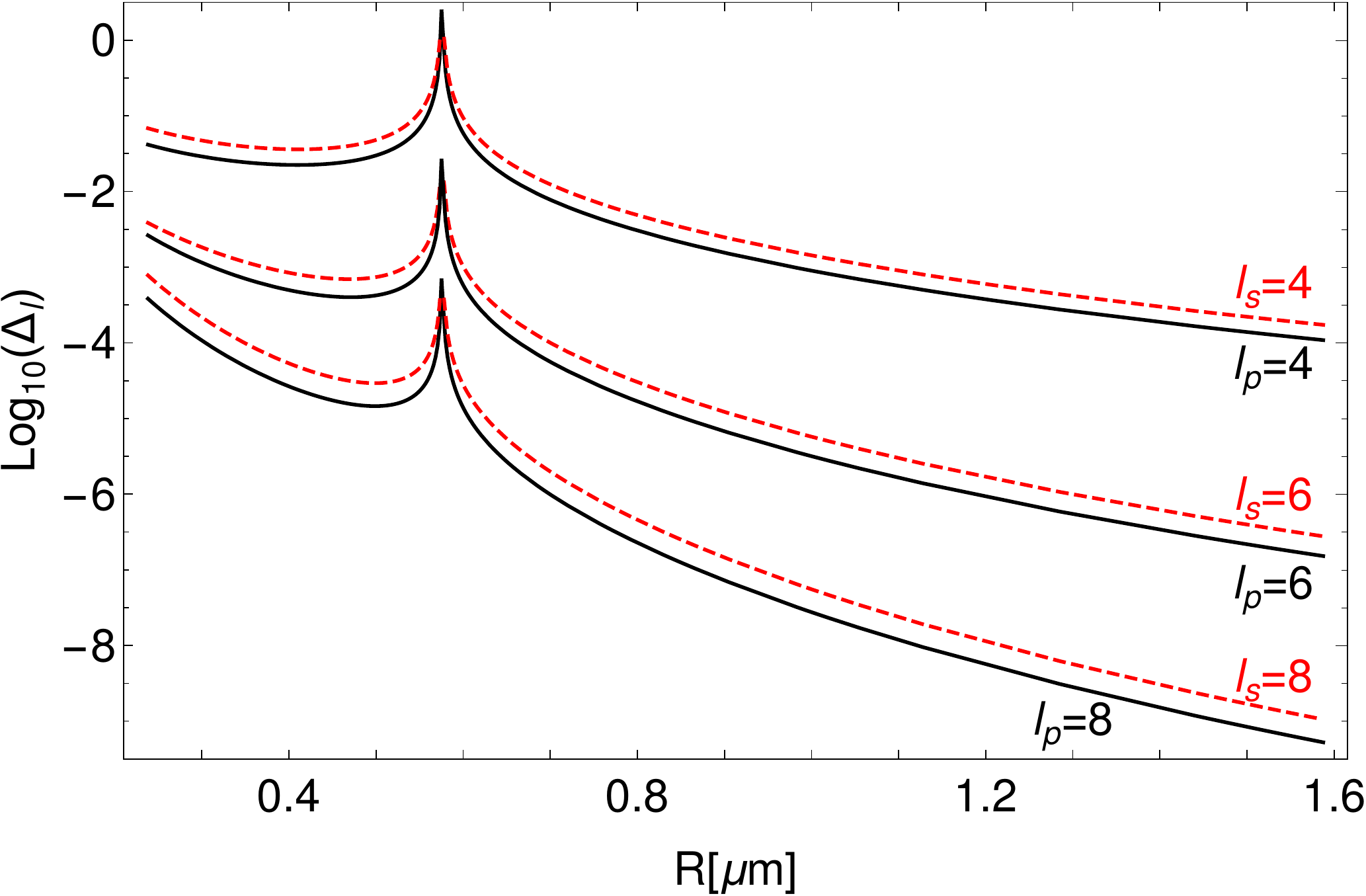}
		\caption{Relative error in the partial sums with respect to the limiting result with $\ell =10$, Eq.~(\ref{eq:er}). The spherical multipole expansion corresponds to the red dotted line. The prolate spheroidal multipole expansion corresponds to the black continuos line. The results correspond to a pair of trilobite $n=33$ molecules with $\mathsf{r}_A = \mathsf{r}_B = 1500\mathrm{a}_0$, parallel natural axes and $\Theta =54.5^\circ$.  }
		\label{fig:rer}
	\end{center}
\end{figure}

The convergence of the spherical and spheroidal expansions was also evaluated by calculating the relative error in the partial sums with respect to the limiting result with $\ell =10$,
\begin{equation}
\Delta_l(R,\Theta)=\left|  \frac{\sum_{\ell\le l}\tilde{\mathcal{V}}_{\ell 0}(R,\Theta)-\sum_{\ell\le 10}\tilde{\mathcal{V}}_{\ell 0}(R,\Theta)}{\sum_{\ell\le 10}\tilde{\mathcal{V}}_{\ell 0}(R,\Theta)}   \right|.\label{eq:er}
\end{equation}
In Fig.~\ref{fig:rer} we report the results for the interesting angle $\Theta =54.5^\circ$. Notice the semilogarithmic scale. For each partial sum the spheroidal multipole results show a faster convergence than the spherical ones.

\subsubsection{Cosine butterfly molecules.}

In this case, the equations determining the intermolecular interaction must incorporate the dependence on the azymuthal angle $\Phi$. It results that 
\begin{equation}
\int \frac{| \Upsilon_A( \vec{r};\vec{r}_A)|^2}{|\vec{r}+\frac{\vec{r}_A}{2}-\vec{R}|} d^3r = \frac{2}{\mathsf{r}_A} \sum_{\ell=0}^{\infty} \sum_{m=-l}^{l} G_{\ell m}(A;\mathsf{r}_A)  \tilde{\mathcal{I}}_{\ell m}^*(\vec{R};\vec{r}_A),
\end{equation}
with
\begin{equation}
G_{\ell m}(A;\vec{\mathsf{r}}_A)= \sum_{j_1,j_2}   t_{m 0 m}^{j_1 j_2 \ell} \frac{j_2!}{(2 j_2-1)!!} \tilde{\mathcal{T}}_{j_1 m}(A;\vec{\mathsf{r}}_A).
\end{equation}
For the cosine butterfly molecules, the multipole moments and the coefficients $G_{\ell m}$ may not be null only if $m=0, \pm 2$. 

\begin{figure}[t]
	\begin{center}
\begin{tabular}{@{}c @{}c}
		\includegraphics[width=.45\textwidth]{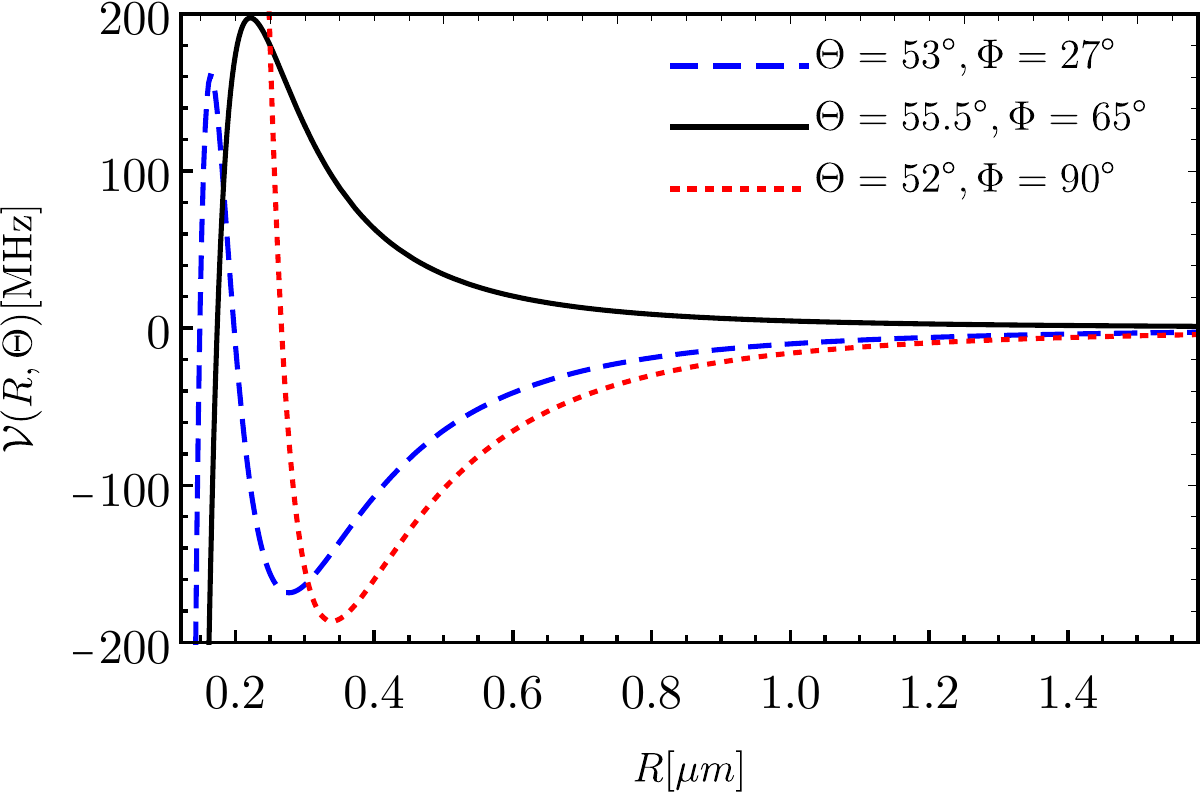}&
                \includegraphics[width=0.45\textwidth]{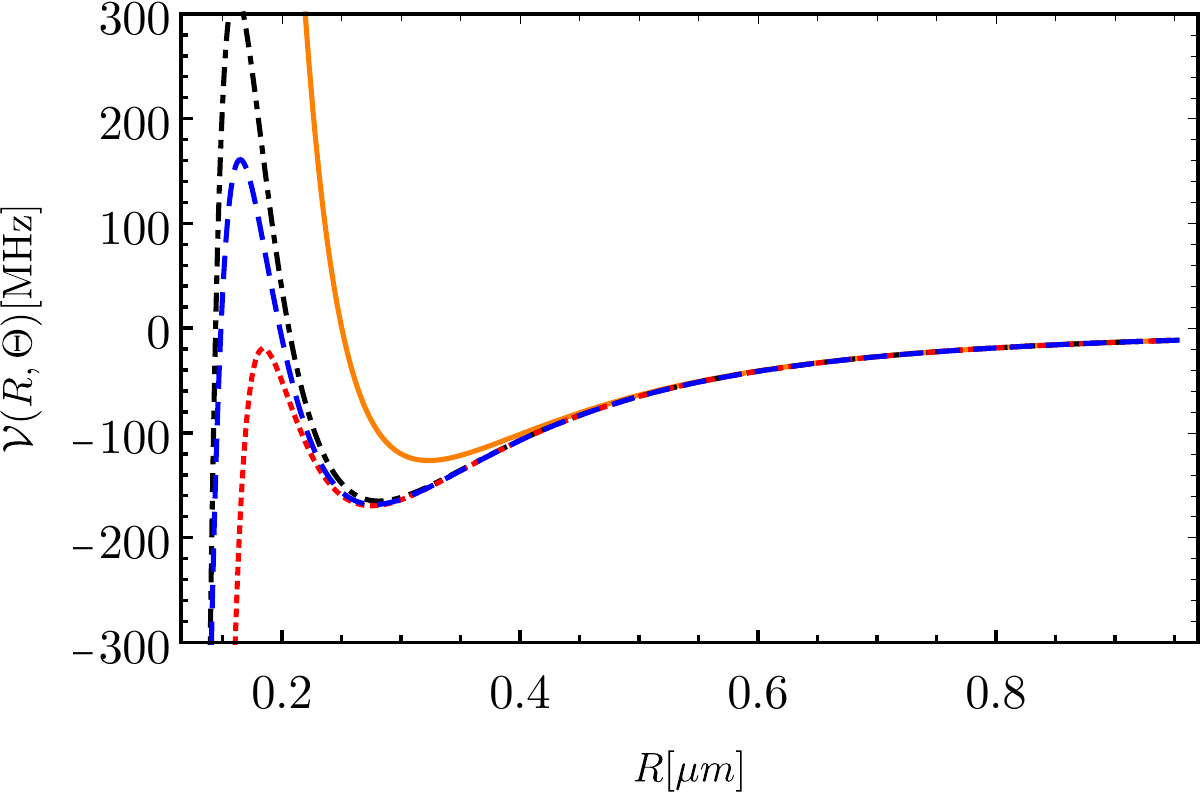}\\
(a) & (b) 		
\end{tabular}
	\end{center}
\caption{(a) Converged intermolecular potential for a pair of cosine butterfly Rb molecules at different configurations. (b)Convergence of the spheroidal multipole expansion for a pair of $n=33$ Rb cosine butterfly molecules in the parallel configuration, $\Theta =53^ \circ$ and $\Phi =27^\circ$; $\ell=4$ corresponds to the orange solid line; $\ell=6$  to black dot-dashed line; $\ell=8$ to red dotted line; $\ell=10$ to blue dashed line. The spherical expansion exhibits a similar convergence.}\label{fig:convcb}

\end{figure}

Assuming $\Omega_A=\Omega_B$, that is parallel axes and parallel planes that define the azimuthal angles, the intermolecular potential becomes 
\begin{align}
\nonumber   \mathcal{V}(\vec{R};A;B)&= \frac{1}{R}+
\frac{2}{\mathsf{r}_A} \sum_{\ell=0}^{\infty} \sum_{m=-\ell}^\ell\left[-G_{\ell m}(A,\mathsf{r}_A)+ \sum_{j_1,m_1} \sum_{j_2,m_2} (-1)^{j_2} t_{m_1 m_2 m}^{j_1 j_2 l} G_{j_1 m_1}(A;\mathsf{r}_A) G_{j_2 m_2}(B;\mathsf{r}_B) \right] \tilde{\mathcal{I}}_{\ell m}^*(\vec{R};\vec{\mathsf{r}}_A) \\
&- \frac{2}{\mathsf{r}_B} \sum_{\ell=0}^{\infty} (-1)^l G_{\ell m}(B;\mathsf{r}_B)  \tilde{\mathcal{I}}_{\ell m}^*(\vec{R};\vec{\mathsf{r}}_B).
\label{eq:intpp}
\end{align}
For two identical cosine butterfly molecules only the $\ell$ even terms contribute to $\mathcal{V}(\vec{R})$
both for the spherical and spheroidal expansions.
For a pair of $n=33$ Rb cosine butterfly molecules with $\mathsf{r}_A=\mathsf{r}_B = 520\mathrm{a}_0$, the partial sums for the total interaction energy  converged within seven  significant figures when up to $\ell=10$ terms were considered.
Figure~\ref{fig:convcb}a illustrates the converged intermolecular potential $\mathcal{V}_{\ell=10}$ for three representative angles. 
The richness of the possible configurations increases with respect to the trilobite case. Nearby the  "magic" angle, at $\Theta = 55.5^\circ$, the blocking phenomenon predicted in Ref.~\cite{eiles2017b} for radial butterfly molecules is observed for a relative azimuthal angle $\Phi =65^\circ$. At slightly lower values of  $\Theta$
both blocking at a distance $R\approx 4\times 10^3\mathrm{a}_0$ followed by a minimum at $R\approx 6\times 10^3\mathrm{a}_0$ occurs for $\Theta\sim 53^\circ$ and $\Phi =27^\circ$.
At an slightly lower value, $\Theta =52^\circ$ a local minimum is predicted for
$\Phi = 90^\circ$. Notice that the energy scale is about ten times higher for the
butterfly than for the trilobite molecules taken in these illustrative examples.
This is due to the smaller natural length scales involved in the butterfly LRRM than in the trilobite. Another consequece of this smaller value of $\mathsf{r}_{A,B}$ is that  the spherical and spheroidal multipole expansions are closer to each other than in  the exemplified trilobite LRRM, so that their convergence is similar. 
In both spherical and spheroidal exapansions, the summation up to $\ell=6$ gives a qualitative description of the converged $\ell =10$ result, as illustrated in Fig.~\ref{fig:convcb}b for  $\Theta\sim 53^\circ$ and $\Phi =27^\circ$.

\section{Discussion}
The simple expression of the  binding electron orbital in terms of prolate spheroidal coordinates supports the description of the electrostatic potential in terms of the corresponding multipole expansion. 
In this work we have studied in detail the electrostatic potential of LRRM and the intermolecular interaction of two LRRM. A comparison of the results from a truncated spherical and prolate spheroidal
multipole expansion was performed for  Rydberg molecules of Rb with $n=33$.

For both trilobite and butterfly molecules, it was found that the prolate spheroidal multipole expansion up to a given $\ell$ moment gives an electrostatic potential that is closer, in general, to the exact result as compared to its spherical analogue.  A natural length scale and orientation encoded in the vector $\vec{\mathsf{r}}$ that joins the neutral atom and the Rydberg core, allows a gross and informative description of LRRM using the single logarithmic monopole term. This vector can also be used to define dimensionless multipole moments and potentials, $\tilde{\mathcal{T}}_{\ell m}$ and $\tilde{\mathcal{I}}_{\ell m}$. Such a definition
permits to compare in an unbiass way multipoles with different $\ell$ values.

In the trilobite case,
the length scale $\mathsf{r}$ is large, $\mathsf{r}_\mathrm{t}\sim 1500\mathrm{a}_0$. The electron orbital is concentrated nearby the perturbing neutral atom. Then, the truncated spheroidal multipole expansion results give clearly a more accurate description than their spherical analogue. The equillibrium distance for $n=33$ radial and angular butterfly molecules is about one third that of a trilobite; the supremacy of spheroidal coordinates over spherical ones at long distances is less evident in those cases, but it persists. The binding electron orbital of a cosine butterfly molecule  is not concentrated nearby the perturbing neutral atom. As a consequence, the value of the dimensionless quadrupole moment $\tilde{\mathcal{T}}_{2,m}$ is almost twice that of the dimensionless dipole moment $\tilde{\mathcal{T}}_{1,0}$. This prevents the application of standard Van der Walls theory for the study of  intermolecular interactions for cosine butterfly molecules.
For both trilobite and butterfly molecules, it was found that a general resemblance of their electrostatic potential is achieved even when the single monopole plus dipole terms are included, i.e., the simple expression Eq.~(\ref{eq:far}) describes most general features  both at short distance and at long distances. Nevertheless, $\ell>1$ multipole moments are required for a detailed description of the intermolecular interactions. For cosine butterfly molecules spheroidal quadrupole terms, Eq.~(\ref{eq:far:ab}), must also be included. Notice that the $Q_{\ell m}(\xi)$ functions decay faster than $\xi^{-(\ell +1)}$ for $\xi\gg 1$.

 As $n$ increases, the potential energy curve of a LRRM acquires more local minima at higher values of the stability distance  $\mathsf{r}$. As a consequence, for higher values of the effective principal number  $n$ the spheroidal description  improves its efficiency.  Experimental results for $n\sim 40$ for Rb and Cs atoms have already been reported~\cite{pohl2011}. An open question concerns to the existence of a dynamical variable that embeds the spheroidal symmetry and gives a deeper physical insight of LRRM's.

The intermolecular interaction between a pair of LRRM exhibits a rich structure. It was shown that the multipole analysis  gives always reliable results whenever up to $\ell=4$ terms were included for $n=33$ trilobite LRRM with parallel orientation of
their vector $\vec{\mathsf{r}}$. This was shown when the molecules are far appart allowing a first order perturbative analysis.
Notice that if the molecules were closer, the supremacy of the spheroidal expansion
would be more evident. The numerical simulations also revealed an effective  radius for which the external multipole expansion can be applied that is smaller for spheroidal than for spherical multipole descriptions.

Interesting results that are evident from the spherical analysis can also be translated to spheroidal coordiantes. For instance, the identification of the "magic" angle at which the spherical dipole-dipole external potential vanishes: $P_{2}^0(\cos\Theta_m) =0$,
can also be written for spheroidal coordinates $P_{2}^0(\eta_m) =0$ with an analogous interpretation. Since for large intermolecular distances $\eta\rightarrow \cos\Theta$,  both spheroidal and spherical exterior potential vanish for this orientation whether it is parametrized by $\Theta_m$ or by $\eta_m$. Nearby  $\Theta_m$ (or $\eta_m$)
the intermolecule potential exhibits blocking and antiblocking features, as well as
a combination of both in the case of cosine butterfly molecules.
This means that, as usual, observing the features of a physical system from different
sights always gives a better picture.

Future analysis should consider that the intermolecular interactions between LRRM at shorter distances requires the incorporation of effects on the distortion of the electronic orbitals due to induced multipole terms.
Having in mind recent experimental analysis, these studies should also involve heteronuclear LRRM \cite{peper2021} and ionic Rydberg molecules~ \cite{duspayev2021}.

An important question is whether prolate spheroidal expansions could be useful for 
other kind of molecules. A partial answer can be found in the current literature. 
The logarithmic structure of the external spheroidal expansion resembles the Hamaker formulation of long range Van der Waals interactions between nanoparticles \cite{roth1996}. Hamaker obtained
an effective potential \cite{hamaker1937} by studying the London-Van der Waals interaction between two spherical particles as a function of their diameters and the distance that separates them.
The resulting natural coordinates are prolate spheroidal as he implicitly recognized in his work.  Nevertheless, a theoretical analysis
at depth about the relevance of alternative multipole expansions in the study of nanoparticle interactions is still lacking.

{\bf Acknowledgements.} This work was partially supported by CONACyT LN-314860 and DGAPA IN-103020 .

\newpage

\section{Appendix}

The unnormalized internal $V^{i}_{\ell m}(\vec r;\vec{ \mathsf{r}})$ and external  $V^{e}_{\ell m}(\vec r;\vec{ \mathsf{r}})$ spheroidal solid harmonics are
\begin{equation}
V^{i}_{\ell m}(\vec r;\vec{ \mathsf{r}}) = \mathcal{P}_\ell^m(\xi)P_\ell^m(\eta)e^{im\varphi};\quad\quad V^{e}_{\ell m}(\vec r;\vec{ \mathsf{r}})= Q_\ell^m(\xi)P_\ell^m(\eta)e^{im\varphi}
\end{equation}
when written in terms of the prolate spheroidal coordinates that take the $z$-axis as paralell to $\vec{ \mathsf{r}}$ and their foci separated a distance $\mathsf{r}$.

Their spherical analogues
\begin{equation}
U^{i}_{\ell m}(\vec r;\hat{ \mathsf{r}}) =  r^{\ell} P_\ell^m(\cos\theta)e^{im\varphi};\quad\quad U^{e}_{\ell m}(\vec r;\hat{ \mathsf{r}})= r^{-(\ell+1)} P_\ell^m(\cos\theta)e^{im\varphi}
\end{equation}
are written in terms of the spherical coordinates $(r,\theta,\varphi)$. 

In these equations $P_\ell^m(x)$ ($x\le 1$), $\mathcal{P}_\ell^m(x)$ ($x\ge 1$) and $Q_\ell^m(x)$ represent the associated Legendre polynomials of first and second kind. 

Spheroidal solid harmonics are connected to spherical solid harmonics by the relations \cite{arnaoudov2010},
\begin{eqnarray}
U_{\ell m}^i(\vec r;\hat{\mathsf{r}}) &=&\sum_{\ell^\prime=0}^\ell\Big[
\frac{(\mathsf{r}/2)^\ell(2\ell^\prime +1)(\ell +m)!(\ell^\prime -m)!}{(\ell + \ell^\prime +1)!!(\ell - \ell^\prime)!!(\ell^\prime +m)!}
\Big]\delta_{\ell m}^{\ell^\prime} V^i_{\ell^\prime m}(\vec r;\vec{\mathsf{r}})
\\
U_{\ell m}^e(\vec r;\hat{\mathsf{r}}) &=&\sum_{\ell^\prime=\ell}^\infty\Big[\frac{(-1)^{\frac{\ell^\prime-\ell}{2}} (2\ell^\prime +1)(\ell + \ell^\prime -1)!!(\ell^\prime -m)!}{(\mathsf{r}/2)^{\ell +1}(\ell -m)!(\ell^\prime - \ell)!!(\ell^\prime +m)!}\Big]\delta_{\ell^\prime m}^{\ell} V^e_{\ell^\prime m}(\vec r;\vec{\mathsf{r}})
\\
V_{\ell m}^i(\vec r,\vec{\mathsf{r}}) &=&\sum_{\ell^\prime=0}^\ell\Big[
\frac{(-1)^{\frac{\ell -\ell^\prime}{2}} (\ell + \ell^\prime -1)!!(\ell +m)!}{ (\mathsf{r}/2)^{\ell^\prime}(\ell^\prime +m)(\ell -\ell^\prime)!!(\ell -m)!}
\Big]\delta_{\ell m}^{\ell^\prime} U^i_{\ell^\prime m}(\vec r;\hat{\mathsf{r}})
\\
V_{\ell m}^e(\vec r,\vec{\mathsf{r}}) &=&\sum_{\ell^\prime=\ell}^\infty\Big[
\frac{(-1)^m (\mathsf{r}/2)^{ \ell^\prime+1} (\ell^\prime -m)!(\ell +m)!}{(\ell ^\prime - \ell)!!(\ell + \ell^\prime +1)!!(\ell -m)!}
\Big]\delta_{\ell^\prime m}^{\ell} U^e_{\ell^\prime m}(\vec r;\hat{\mathsf{r}})
\end{eqnarray}
with $\delta_{nk}^l = 1$ if $(n-l)$ is even and $k\le l\le n$, and zero otherwise.

Some relevant expressions related to spheroidal harmonics were
reported in Ref.\cite{jansen2000} and, here, we summarize them.

The similitudes between  spherical and spheroidal harmonics are emphasized by defining the prolate spheroidal functions,
\begin{eqnarray}
{\mathcal{R}}_{\ell m} (\vec r;\vec{\mathsf{r}}) &=&\Big(\frac{\mathsf{r}}{2}\Big)^\ell\frac{(\ell -m)!}{(2\ell -1)!!}
\sqrt{\frac{(\ell -m)!}{(\ell+m)!}} V^i_{\ell m}(\vec r;\vec{\mathsf{r}}),\\
\mathcal{I}_{\ell m} (\vec r;\vec{\mathsf{r}}) &=&\Big(\frac{\mathsf{r}}{2}\Big)^{-\ell-1}\frac{(2\ell+1)!!}{(\ell +m)!}
\sqrt{\frac{(\ell -m)!}{(\ell+m)!}} V^e_{\ell m}(\vec r;\vec{\mathsf{r}}).
\end{eqnarray}
They have as asymptotic limit for $\mathsf{r}\rightarrow 0$ their spherical analogues ~\cite{jansen2000}.
\begin{eqnarray}
\mathcal{Q}_{\ell m} (\vec r) &=&\sqrt{\frac{(\ell -m)!}{(\ell +m)!}} U^i_{\ell m}(\vec r),\\
\mathcal{J}_{\ell m} (\vec r) &=&\sqrt{\frac{(\ell -m)!}{(\ell +m)!}}  U^e_{\ell m}(\vec r).
\end{eqnarray}
${\mathcal{R}}_{\ell m}$ has a compact expression under traslation, rotation and scaling transformations~\cite{jansen2000}. For translations,
\begin{eqnarray}
{\mathcal{R}}_{\ell m} (\vec a +\vec b;\vec{\mathsf{r}}_A) &=&\sum_{j_1j_2} \sum_{m_1m_2}\big(\frac{\mathsf{r}_A}{2}\Big)^{\ell -j_1-j_2} \;t_{m_1m_2 m}^{j_1 j_2 \ell}{\mathcal{R}}_{j_1 m_1} (\vec a ;\vec{\mathsf{r}}_A){\mathcal{R}}_{j_2 m_2} (\vec b;\vec{\mathsf{r}}_A)\label{eq:translation}\\
t_{m_1m_2 m}^{j_1 j_2 \ell}& =&\tilde \Delta_{m_1m_2 m}^{j_1 j_2 \ell} \frac{(2j_1+1)!!(2j_2+1)!!}{(2\ell -1)!!}
\sqrt{\frac{(\ell+m)!(\ell-m)!}{(j_1+m_1)!(j_1-m_1)!(j_2+m_2)!(j_2-m_2)!}}\nonumber\\
&\times& \sum_{i_1 = j_1}^\prime\sum_{i_2 = j_2}^\prime\frac{(-1)^{(\ell-i_1-i_2)/2}(\ell +i_1+i_2-1)!!}{(\ell -i_1 -i_2)!!(i_1 -j_1)!!(i_1+j_1+1)!!(i_2 -j_2)!!(i_2+j_2+1)!!} \label{eq:primed}\\
\tilde \Delta_{m_1m_2 m}^{j_1 j_2 \ell}&=&    \begin{cases} 1& (j_1+j_2+\ell)\; {\mathrm{even}},\;\; m_1+m_2=m\\
                                                              & \ell\ge j_1+j_2\ge\vert m\vert \;\; j_1\ge \vert m_1\vert, \;\; j_2\ge \vert m_2\vert \\ 0, &\text{otherwise}  \end{cases}.     
\end{eqnarray}
The prime at the summmation sign on Eq.~\ref{eq:primed} indicates that $i_1+i_2\le \ell$,  and $i_1$ and $i_2$ vary in steps of two.

If a rotation defined by the Euler angles $\Omega=(\alpha,\beta,\gamma)$ is performed,
\begin{eqnarray}
{\mathcal{R}}_{\ell m} (\vec r^\prime;\vec{\mathsf{r}}_A) &=&\sum_{k} \sum_{m^\prime} \Big(\frac{\mathsf{r}_A}{2}\Big)^{\ell -k} \mathcal{D}_{m^\prime m}^{k \ell} (\Omega)
{\mathcal{R}}_{k m^\prime} (\vec r ;\vec{\mathsf{r}}_A) \label{eq:rotation}\\
\mathcal{D}_{m^\prime m}^{k \ell} (\Omega)& =& e^{-i(\alpha m^\prime + \gamma m)}{d}_{m^\prime m}^{k \ell} (\beta)\nonumber \\
{d}_{m^\prime m}^{k \ell} (\beta)& = &\Delta_{m}^{\ell k} \Delta_{m^\prime}^{k k}\frac{(2k+1)!!}{(2\ell -1)!!}
\sqrt{\frac{(\ell+m)!(\ell-m)!}{(k+m^\prime)!(k-m^\prime)!}}{\sum_{i= I_1}^\ell}^\prime\frac{(-1)^{(\ell -i)/2}(\ell +i -1)!!}{(\ell -i)!!(i-k)!!(i+k+1)!!}\nonumber\\
&\times& \sum_{i^\prime = S_1}^{S_2} (-1)^{i^\prime}\begin{pmatrix} i + m^\prime\\ i^\prime\end{pmatrix}
\begin{pmatrix} i - m^\prime\\ i^\prime + m -m^\prime\end{pmatrix}
(\cos\beta/2)^{2i+m^\prime-m -2i^\prime}(\sin\beta/2)^{2i^\prime+m^\prime-m},\label{eq:des}\\
\Delta^{\ell k}_m &=& \begin{cases} 1 &(\ell-k)\text{even},\ \ \ell\ge k\ge \vert m\vert\\ 0, &\text{otherwise}  \end{cases}.
\end{eqnarray}
Here $I_1$ = max$(k,\vert m\vert)$ for $(\ell -m)$ even and $I_1$ = max$(k,\vert m\vert +1)$ else,
while the sum over $i^\prime$ runs from $S_1=$ max$(0, m^\prime - m)$ to $S_2=$ max$(i +m^\prime, i-m)$. The coefficients $\mathcal{D}_{m^\prime m}^{k \ell}(\Omega)$ are the spheroidal analogues of Wigner matrices. The prime at the summmation sign on Eq.~\ref{eq:des} indicates that the index varies in steps of two. 

For a scaling transformation of otherwise parallel vectors $\vec{\mathsf{r}}_A$ and $\vec{\mathsf{r}}_B$
$$
{\mathcal{R}}_{\ell m}(\vec r;\vec{\mathsf{r}}_A)=\sum_k \Big(\frac{\mathsf{r}_A}{2}\Big)^{\ell -k}\;\mathcal{S}_m^{\ell k}(\mathsf{r}_B/\mathsf{r}_A){\mathcal{R}}_{k m}(\vec r;\vec{\mathsf{r}}_B)$$\begin{equation}
\mathcal{S}_m^{\ell k}(x)=\Delta^{\ell k}_m\Big(\delta_{\ell k} + (1-\delta_{\ell k})
\frac{(-1)^{(\ell-k)/2}}{\ell -k}
\sqrt{\frac{(\ell +m)!(\ell -m)!}{(k+m)!(k-m)!}}\frac{(2k+1)!!}{(2\ell -1)!!}
\sum_{j=0}^{(\ell -k)/2}(-1)^j \begin{pmatrix} \frac{\ell -k}{2} \\ j \end{pmatrix}\begin{pmatrix} \frac{\ell +k -1}{2} +j \\ \frac{\ell -k}{2} -1 \end{pmatrix} x^{2j}\Big)
\end{equation}

The coefficients $\Delta^{\ell k}_m $ and $\tilde \Delta_{m_1m_2 m}^{j_1 j_2 \ell}$ extend the spherical triangle restrictions to the spheroidal domain.

\bibliographystyle{unsrt}
\bibliography{referencias.bib}

\end{document}